\def\year{2021}\relax
\documentclass[letterpaper]{article} % DO NOT CHANGE THIS
\usepackage{aaai21}  % DO NOT CHANGE THIS
\usepackage{times}  % DO NOT CHANGE THIS
\usepackage{helvet} % DO NOT CHANGE THIS
\usepackage{courier}  % DO NOT CHANGE THIS
\usepackage[hyphens]{url}  % DO NOT CHANGE THIS
\usepackage{graphicx} % DO NOT CHANGE THIS
\usepackage{natbib}  % DO NOT CHANGE THIS AND DO NOT ADD ANY OPTIONS TO IT
\usepackage{caption} % DO NOT CHANGE THIS AND DO NOT ADD ANY OPTIONS TO IT
\usepackage{multirow}
\usepackage{booktabs}       % professional-quality tables
\usepackage{amsfonts}       % blackboard math symbols
\usepackage{nicefrac}       % compact symbols for 1/2, etc.
\usepackage{microtype}      % microtypography
\usepackage{lipsum}
\usepackage{romannum}
\usepackage{amsmath}
\usepackage{amssymb}
\usepackage{mathtools}
\usepackage{bm}
\usepackage{cleveref}% for reference \cref
\usepackage{graphicx}
\usepackage{comment}
\DeclareMathOperator*{\E}{\mathbb{E}}
\newcommand{\rpm}{\sbox0{$1$}\sbox2{$\scriptstyle\pm$}
  \raise\dimexpr(\ht0-\ht2)/2\relax\box2 }
  
\usepackage{subcaption}
\usepackage{tikz}
\usepackage{amsmath}
\usetikzlibrary{bayesnet}

\title{On Estimating Recommendation Evaluation Metrics under Sampling}
\author {
        Ruoming Jin,\textsuperscript{\rm 1}\ \ \ 
        Dong Li, \textsuperscript{\rm 1}\ \ \ 
        Benjamin Mudrak\textsuperscript{\rm 1} \ \ \
        Jing Gao, \textsuperscript{\rm 2}\ \ \ 
        Zhi Liu\textsuperscript{\rm 2} \\
}
\affiliations {
    \textsuperscript{\rm 1} Kent State University\ \ \ \ 
    \textsuperscript{\rm 2} iLambda \\
    \{rjin1,dli12,bmudrak1\}@kent.edu\ \ \ \  
    \{jgao,zliu\}@ilambda.com
}
\begin{document}

\maketitle

\begin{abstract}
Since the recent study ~\cite{Krichene20@KDD20} done by Krichene and Rendle on the sampling-based top-k evaluation metric for recommendation, there has been a lot of debates on the validity of using sampling to evaluate recommendation algorithms. Though their work and the recent work ~\cite{Li@KDD20} have proposed some basic approaches for mapping the sampling-based metrics to their global counterparts which rank the entire set of items, there is still a lack of understanding and consensus on how sampling should be used for recommendation evaluation. 
The proposed approaches either are rather uninformative (linking sampling to metric evaluation) or can only work on simple metrics, such as Recall/Precision~\cite{Krichene20@KDD20,Li@KDD20}. 
In this paper, we introduce a new research problem on learning the empirical rank distribution, and a new approach based on the estimated rank distribution, to estimate the top-k metrics. Since this question is closely related to the underlying mechanism of sampling for recommendation, tackling it can help better understand the power of sampling and can help resolve the questions of if and how should we use sampling for evaluating recommendation. 
We introduce two approaches based on MLE (Maximal Likelihood Estimation) and its weighted variants, and ME (Maximal Entropy) principals to recover the empirical rank distribution, and then utilize them for metrics estimation. The experimental results show the advantages of using the new approaches for evaluating recommendation algorithms based on top-k metrics.
\end{abstract}

\section{Introduction}

Recommendation and personalization continue to  play important roles in the deep learning area. Recent studies report that in big enterprises, such as Facebook, Google, Alibaba, etc., deep learning- based recommendation takes the majority of the AI-inference cycle in their production cloud~\cite{deeprec}. However, several recent studies ~\cite{RecSys19Evaluation, Steffen@19DBLP} have called the validity of some recent (mostly deep learning-based) recommendation results into question, particularly highlighting the ad-hoc nature of evaluation protocols, including selective (likely weak) baselines and evaluation metrics. Those factors may lead to false signals of improvements. 

One of the latest controversies comes from the validity of using sampling for evaluating recommendation models: Instead of ranking all available items (whose number can be very large) for each user, a fairly common practice in academics, as well as in industry, is to sample a smaller set of (irrelevant) items, and rank the relevant items against the sampled items~\cite{Koren08, CremonesiKT@10,he2017neural, ebesu2018collaborative,HuSWY18,krichene2018efficient,wang2018explainable,YangBGHE18,YangCXWB18}. Rendel ~\cite{rendle2019evaluation} together with Krichene ~\cite{Krichene20@KDD20} argued that commonly used (top-$k$) evaluation metrics, such as Recall (Hit-Ratio)/Precision, Average Precision (AP) and NDCG, (other than AUC), are all ``inconsistent'' with respect to the global metrics (even in expectation). They suggest the cautionary use (avoiding if possible) of the sampled metrics for recommendation evaluation, and they also propose a few approaches to help correct the sampled metrics to be closer to their global counterparts. 

%Intuitively their fix is to replace the original metric at each sampled rank $r$, denoted as $M(r)$ (for example, for $AP$, $M(r)=1/r$), with a corrected value: $\overhat{M}(r)$. 

In the meantime, the latest work by Li et. al. ~\cite{Li@KDD20} studies the problem of aligning sampling top-$k$ ($SHR@k$) and global top-$K$ ($HR@K$) Hit-Ratios (Recalls) through a mapping function $f$ (mapping the $k$ in the sampling to the global top $f(k)$), so that $SHR@k\approx HR@f(k)$. Basically, the sampling- based top $k$ Hit-Ratio, $SHR@k$, corresponds to the global top-$f(k)$ Hit-Ratio.  They develop methods to approximate the function $f$, and they show that it is approximately linear (the ``sampling'' location of the global top-$K$ curve is almost equally intervaled). However, their methods are limited to only the Recall/Hit-Ratio metric and cannot be generalized to more complex metrics, such as AP and NDCG. 

Despite these latest works~\cite{Li@KDD20, Krichene20@KDD20}, the very question as to if and how sampling can be used for recommendation evaluation remains unsolved and under heavy debate. 
The proposed approaches to estimate the global evaluation metrics based on sampling either are rather uninformative ~\cite{Krichene20@KDD20} or can only work on simple metrics~\cite{Li@KDD20}.  They also provide little insight into how sampled recommendation ranking results can relate to their global counterparts. Particularly, even though methods such as MLE and/or Bayesian approaches are widely used for sampling-based parameter and distribution inference in statistics~\cite{theorypoint}, it remains an open problem if and how they can be leveraged to develop sampling-based estimators for recommendation evaluation metrics. 

\subsection{Contributions and Organization}
%In ~\ref{Problem Formulation},  we formally propose the problem.

%In ~\ref{ch:estimators}, we propose various statistical models to estimate the empirical evaluation metric. 

To address these questions, we make the following contributions in this paper: 

\begin{itemize}
    \item  We introduce a new research problem on learning the empirical rank distribution and a new metric estimation framework based on the learned rank distribution. This estimation framework can allow us to handle all the existing metrics in a unified and more informative fashion. It can be considered as being metric-independent: once the empirical rank distribution is learned, it can be used immediately to estimate any top-$K$ metrics.   
    \item  We introduce two types of approaches for estimating the rank distribution.  The first approach is based on (weighted) MLE, and the second approach is based on combining maximal entropy with a distribution difference constraint. 
    \item  We perform a thorough experimental evaluation on the proposed new estimators for recommendation metrics.  The experimental results show the advantages of using our approaches for evaluating recommendation algorithms based on top-k metrics against the existing ones in ~\cite{Krichene20@KDD20}. 
    \end{itemize}

Our results provide further evidence that sampling can be used for recommendation evaluation. They also further confirm what was first discovered in ~\cite{rendle2019evaluation}: The metrics such as NDCG and AP should not be directly evaluated on top of the sampling rank distributions. More importantly, our results further clarify that those metrics should be applied on the learned empirical rank distribution based on sampling.

\section{Evaluation Metrics and Notation}

\begin{scriptsize}
      \begin{table}
      \begin{minipage}{0.48\textwidth}
      \caption{Notations}
      \vspace{-2.0ex}
  \label{tab:notation}
 \begin{tabular}{|l|l|}
 %\hline
 %\noindent
\hline
$M$ & \# of users in testing data \\
\hline
$N$: & \# of items \\
\hline
$I$	& entire set of items, and $|I| = N$ \\
\hline
$R$	& item rank in range $[1, N]$ \\
\hline
$i_u$	& relevant item for user $u$ in testing data\\
\hline 
$R_u$	& rank of item $i_u$ among $I$ for user $u$	\\
\hline
$n$ & $n-1 = $ \# of sampled items for each user \\
\hline
$I_u$	& $I_u \backslash i_u$ consists of $n-1$ sampled items for $u$ \\
\hline
$r$ &  item rank in range $[1, n]$ \\
\hline
$r_u$	& rank of item $i_u$ among $I_u$\\
%$x_u$   & $x_u=r_u-1$: \# of sampled items ranked  $i_u$\\
\hline 
%$r_R$	& $:= \{r_u: R_u = R\}$ \\
%\hline 
$\mathcal{R}$ & discrete random variable ($\mathcal{R}: u \rightarrow R_u$) \\
\hline
$\pi_R$ & $= Pr(\mathcal{R}=R)$, rank distribution (pmf of $\mathcal{R}$) \\
\hline 
$P(R)$ & empirical rank distribution (pmf) \\
\hline
% $p_u$ & $Pr(r_u\le k)$, probability that $r_u \le k$ among $I_n$ for a user $u$\\
%     \hline
%     $p^R$ & $:= p^u$, for the group of users whose $R_u = R$\\
%     \hline
%     $p_u$&$:= \frac{R_u - 1}{N-1}$, probability for sampling an item that ranks higher than $i_u$ for $u$\\
%     \hline
  \end{tabular}
  \end{minipage}
  \end{table}
\end{scriptsize}

% $I$: total item set

% $I_n$: user specific, $n-1$ random sampled irrelevant items together with $i_u$

% $x_u$: the number of sampled items that are ranked in front of the relevant item $i_u$ for user $u$

% $r_u$: rank of $i_u$ among $I_n$ for user $u$, $r_u = x_u + 1$

% $R_u$: rank of $i_u$ among $I$ for user $u$

% $R$: integer variable, referring to item rank position, in range $[1, N]$

% $r$: integer variable, referring to item rank position, in range $[1, n]$

% $\mathcal{R}$: a random variable, measurable function, mapping users to their relevant items rank among total items. $\mathcal{R}: u \rightarrow R_u$

% $\pi_R$: $= P(\mathcal{R}=R)$,probability mass function of $\mathcal{R}$, probability of users whose relevant items rank is $R$

% ${\mathcal M}$: metric

% $\hat{*}$ : symbol to notate estimation. For example, $\hat{{\mathcal M}}$, estimator of ${\mathcal M}$  

% $P$: denote probability 

% \section{Problem Formulation}\label{ch:problem}

 %Assume the entire dataset is split into training and testing. 
 In this paper, we are mainly concerned with the evaluation of recommendation algorithms in the testing dataset, whose key notations are listed in Table ~\ref{tab:notation}. 
 Given a user $u$ and a (relevant) item $i_u$, the recommendation algorithm $A$ returns $R_u$, the rank of item $i_u$ among all items in set $I$: $R_u=A(u, i_u; I)$.

% Assume that the entire dataset is split into training and testing. 
% Let the testing dataset consist of $M$ users and $N=|I|$ items, where $I$ is the entire set of items. From the training, we will learn a recommendation algorithm $A$, which can rank a given set of items for a user.  To simplify our discussion, we consider the \textit{leave-one-out} strategy ~\cite{mmanuelXBS@16}, where {\bf each user has and only has one relevant item to be evaluated}, though such treatment can be naturally generalized to the situation where a user may have more than one targeted item in the testing data~\cite{ElkahkySH@WWW'15,liang2018variational}.  

Let $\mathcal{M}_{metric}$ be a function (metric) which weighs the relevance or importance of rank position $R$. Then the metric ($metric$) for evaluating the performance of a recommendation algorithm $A$ is simply the average of the weight function:
\begin{small}
\begin{equation}\label{eq:nonconstraint}
metric=\frac{1}{M}\sum_{u=1}^M {\mathcal M}_{metric}(R_u)=\frac{1}{M}\sum_{u=1}^M {\mathcal M}_{metric}(A(u,i_u;I)) 
\end{equation}
\end{small}

The commonly used ${\mathcal M}_{metric}$ for evaluation metrics~\cite{Krichene20@KDD20}, (AUC, NDCG, and AP) are: ${\mathcal M}_{AUC}(R) =\frac{N-R}{N-1};$
\begin{equation}
{\mathcal M}_{NDCG}(R)= \frac{1}{\log_2(R+1)};\ \ \
{\mathcal M}_{AP}(R) =\frac{1}{R}  \nonumber
\end{equation}

Given this, each metric can be defined accordingly. For instance, we have: 
\begin{equation}
AP=\frac{1}{M}\sum_{u=1}^M {\mathcal M}_{AP}(R_u)=\frac{1}{M} \sum_{u=1}^M \frac{1}{R_u} \nonumber
\end{equation}

\subsection{Top-K Evaluation Metrics}
For most of the recommendation applications, only the top-ranked items are of interest. Thus, the commonly used evaluation metrics are primarily based on top-k partial ranked lists. Specifically, the corresponding weight/importance of the relevant item $i_u$ will only be counted in the overall metrics if $i_u$ is ranked higher than $k$. Mathematically, the weight function ${\mathcal M}^K$ (for top-K evaluation) will include an indicator term (${\bf 1}_{X}=1$ iff $X$ is true, and $0$ otherwise): 
\begin{equation}
    {\mathcal M}^K_{metric}(R)={\bf 1}_{R\leq K} {\mathcal M}_{metric}(R)
\end{equation}
where {\em metric} includes the aforementioned methods such as AUC, NDCG, and AP, as well as the commonly used Recall (Hit-Ratio) and Precision, whose importance metrics are constant:
\begin{equation}
{\mathcal M}_{Recall}(R) = 1; \ \ \ {\mathcal M}_{Prec}(R)=1/K  \nonumber
\end{equation}

Given this,  the top-K evaluation metrics,  $metric@K=$ 
\begin{small}
\begin{equation}
\frac{1}{M}\sum_{u=1}^M {\mathcal M}^K_{metric}(R_u)=\frac{1}{M}\sum_{u=1}^M {\bf 1}_{R_u\leq K} {\mathcal M}_{metric}(R_u) 
\end{equation}
\end{small}

The commonly used top-K metrics include Recall@K, Precision@K, AUC@K, NDCG@K and AP@K, among others. For instance, 
\begin{equation}
AP@K=\frac{1}{M}\sum_{u=1}^M {\mathcal M}^K_{AP}(R_u)=\frac{1}{M} \sum_{u=1}^M {\bf 1}_{R_u\leq K} \frac{1}{R_u} \nonumber
\end{equation}
We note the unconstrained metrics defined in Equation~\ref{eq:nonconstraint} are the special case of top-K metrics, where $K=N$. Thus, we will focus on studying the top-K evaluation metrics. 

% Note that as $AUC$ can be estimated without bias (unbiased estimator) by sampling and is deemed ``consistent''~\cite{}, we will not consider it here.  
% Empirically, the equation that has been widely taken to evaluate the
% performance of a recommender algorithm is:

% \begin{equation}\label{eq:empirical_metric}
%      \E{[{\mathcal M}(R)]}=\sum\limits_{R=1}^{N}{{\mathcal M}(R)}P(R)
%     \approx \frac{1}{M}\sum\limits_{u=1}^{M}{{\mathcal M}(R_u)}
% \end{equation}
% where ${\mathcal M}$ is the metric function, typical top-K metrics, like AP@K, Recall@K, NDCG@K, etc.

% Formally, given the observations $\{ x_u \}_{u=1}^{M}$ (or $\{ r_u \}_{u=1}^{M}$) under sampling, the problem is to estimating \cref{eq:empirical_metric} without knowledge of $\{ R_u \}_{u=1}^{M}$.

\subsection{Sampling Top-K Evaluation}
Under the sampling-based top-K evaluation, 
for a given user $u$ and his/her relevant item $i_u$, only $n - 1$ irrelevant items from the entire set of items $I$ are sampled, together with $i_u$ forming $I_u$ ($i_u \in I_u$, $|I_u|=n$). 
Thus, the rank of $i_u$ among $I_u$ is denoted as $r_u = A(u, i_u; I_u)$.

Given this, a (seemingly) natural and also commonly used practice in recommendation studies ~\cite{Koren08, he2017neural,liang2018variational} is to simply replace $R_u$ with $r_u$ for (top-$K$) evaluation, denoted as   
% \begin{equation}
% \frac{1}{M}\sum_{u=1}^M {\mathcal M}(r_u)=\frac{1}{M}\sum_{u=1}^M {\mathcal M}(A(u,i_u;I_u)) 
% \end{equation}

% Accordingly, the top-K evaluation metric  
$\widetilde{metric}@K=$
\begin{equation}\label{eq: empirical_m}
\frac{1}{M}\sum_{u=1}^M {\mathcal M}^K_{metric}(r_u)=\frac{1}{M}\sum_{u=1}^M {\bf 1}_{r_u\leq K} {\mathcal M}_{metric}(r_u) 
\end{equation}
For instance, the sampling top-K AP metric is
\begin{equation}
\widetilde{AP}@K=\frac{1}{M}\sum_{u=1}^M {\mathcal M}^K_{AP}(r_u)=\frac{1}{M} \sum_{u=1}^M {\bf 1}_{r_u\leq K} \frac{1}{r_u} \nonumber
\end{equation}

\subsection{The Problem of $\widetilde{metric}@K$}
It is rather easy to see that the range of sampling rank $r_u$ (from $1$ to $n$) is very different from the range of true rank $R_u$ (from $1$ to $N$) of any user $u$. Thus, for the same $K$, the sampling top-K metrics and the global top-K correspond to very different measures (no direct relationship): 
\begin{equation}
metrics@K  \neq \widetilde{metrics}@K 
\end{equation}
This is the ``problem'' being highlighted and confirmed in~\cite{Krichene20@KDD20, rendle2019evaluation}, and they further formalize that these two metrics are ``inconsistent''. 
Using statistics terminology, the commonly used sampling-based top-$K$ metric $\widetilde{metric}@K$  is not a ``reasonable'' {\em estimator} ~\cite{theorypoint} of the exact $metrics@K$ from the entire testing data. 

However, Li et. al. ~\cite{Li@KDD20} showed that for some of the most commonly used metrics, the Recall/HitRatio, there is a  mapping function $f$ (approximately linear), such that 
\begin{equation}
Recall@f(K) \approx \widetilde{Recall}@K 
\end{equation}
In other words, for Recall at $f(1)$, $f(2)$, $\dots$, $f(n)=N$, they can be estimated by the the sampling-based top-$K$ Recall/HitRatio $\widetilde{Recall}$ at with $K=1$, $K=2$, $\cdots$, $K=n$, respectively. 
Note that this result can be generalized to the Precision metrics, but it has difficulty for more complex metrics, such as NDCG and AP~\cite{Li@KDD20}.

\section{Top-$K$ Metrics Estimation}

Now, we formally introduce the estimation problem of the (top-K) evaluation metrics under sampling. 
Given the sampling ranked results in the testing dataset, $\{r_u\}^{M}_{u=1}$, we would like to develop various estimators $\widehat{metric}@K$ to approximate $metric@K$ (Equations~\ref{eq: empirical_m}), i.e. 
\begin{equation}
    metric@K \approx \widehat{metric}@K
\end{equation}
Note that in general, we would like the estimators to have low bias and variance (or be unbiased), among other desirable properties~\cite{theorypoint}. 

%in statistics, statistical properties, such as bias, variance, consistency, and efficiency, etc. to theoretically quantify the estimators. 

\subsection{The Sampled Metric $\widehat{\mathcal M}(r)$ Approach}
In ~\cite{Krichene20@KDD20}, Krichene and Rendle notice that the overall metrics ($metric@K$) are the average of the weighting function (${\mathcal M}^K_{metric}(R_u)={\bf 1}_{R\leq K} {\mathcal M}_{metric}(R)$).
Their approach is to develop a {\em sampled metric} $\widehat{\mathcal M}^K_{metric}(r)$ ($\widehat{\mathcal M}(r)$ for simplicity) so that: 
\begin{small}
\begin{equation}
    \frac{1}{M}\sum_{u=1}^M {\mathcal M}^K_{metric}(R_u) \approx \frac{1}{M}\sum_{u=1}^M \widehat{\mathcal M}(r_u)  \ \ \ \Big(= \sum_{r=1}^n \tilde{P}(r) \widehat{\mathcal M}(r)  \Big)
\end{equation}
\end{small}
where $\tilde{P}(r)=\frac{1}{M} \sum_{u=1}^M {\bf 1}_{r_u=r}$ is the empirical rank distribution on the sampling data. 

They have proposed a few estimators based on this idea, including estimators that use the unbiased rank estimators,  minimize bias with monotonicity constraint ($CLS$), and utilize Bias-Variance ($BV$) tradeoff. Their study shows that only the last one ($BV$) is competitive ~\cite{Krichene20@KDD20}. We describe it below. 

\subsubsection{Bias-Variance ($BV$) Estimator}
The $BV$ estimator is to consider the tradeoff between two goals: 1) minimize the difference between $metric@K$ and the expectation of the estimator, which can be written as 
\begin{equation}
E\Big(\frac{1}{M}\sum_{u=1}^M \widehat{\mathcal M}(r_u)\Big) = \frac{1}{M}\sum_{u=1}^M E\Big(\widehat{\mathcal M}(r_u)|R_u\Big)
\end{equation}
and 2) minimize the sum of variance of $\widehat{\mathcal M}(r_u)$ given its global rank $R_u$, $\sum_{u=1}^M Var[\widehat{\mathcal{M}}(r)|R]$. 
Let $P(R)$ be the empirical pmf (probability mass function) for the rank distribution $
    P(R)=\frac{1}{M} \sum_{u=1}^M {\bf 1}_{R_u=R}$. 
Then, the $BV$ estimator uses the $n$ dimensional vector $\widehat{\mathcal M}:=(\widehat{\mathcal M}(r))_{r=1}^n \in \mathbb{R}^n$ to minimize the following formula:
\begin{small}
\begin{equation}
   \sum\limits_{R=1}^{N}{P(R)
   \Big((\E{\big[\widehat{\mathcal{M}}(r)|R\big]-\mathcal{M}^K_{metric}(R))^2  
   +\gamma \text{Var}[\widehat{\mathcal{M}}(r)|R ] }\Big)}
\end{equation}
\end{small}

Since this is a regularized least squares problem, its optimal solution is ~\cite{Krichene20@KDD20}:
\begin{equation} \label{eq:leastsquare}
\widehat{\mathcal{M}} = \Big((1.0-\gamma)A^TA+\gamma \text{diag}(\pmb{c})  \Big)^{-1}A^T\pmb{b}
\end{equation}
where 
\begin{equation}\label{eq:BV_param}
    \begin{split}
        &A\in \mathbb{R}^{N\times n},\quad A_{R,r} = \sqrt{P(R)}P(r|R)\\
        &\pmb{b}\in \mathbb{R}^N,\quad b_{R} = \sqrt{P(R)}\mathcal{M}^K_{metric}(R)\\
        &\pmb{c}\in \mathbb{R}^n, \quad c_{r} = \sum\limits_R^{N}{P(R)P(r|R)}
    \end{split}
\end{equation}
Since the rank distribution $P(R)$ is unknown, they simply use the uniform distribution in ~\cite{Krichene20@KDD20} and found it works reasonably well. 
Furthermore, they empirically found that when $\gamma\leq 0.1$ they achieve a good estimation. 
%Finally, we note that to evaluate any individual metrics $metric@K$ (and $K$ in top-$K$), $BV$ has to perform a separate least square optimization procedure to identify its corresponding $\widehat{\mathcal M}^K_{metric}$ ($\widehat{\mathcal M}$). 

\subsection{The New Approach and New Problem}
Our new approach is based on the following observation: 
\begin{small}
\begin{equation}
metric@K=\frac{1}{M}\sum_{u=1}^M {\mathcal M}^K_{metric}(R_u)=\sum_{R=1}^K P(R) {\mathcal M}_{metric}(R) 
\end{equation}
\end{small}
Thus, if we can estimate $\widehat{P}(R)\approx P(R)$, then we can derive the metric estimator as 
\begin{equation}\label{eq:metric}
\widehat{metric}@K=\sum_{R=1}^K \widehat{P}(R) {\mathcal M}_{metric}(R)
\end{equation}

\subsubsection{New Problem} Given this, we introduce the problem of learning the empirical rank distribution $(P(R))_{R=1}^N$ based on sampling $\{r_u\}_{r=1}^M$.  In general, only when $R$ is small is $P(R)$ of interest for estimating the top-$K$ metrics. To our best knowledge, this  problem has not been formally and explicitly studied before for sampling-based recommendation evaluation. 

We note that the importance of the problem is two-fold. 
On one side, the learned empirical rank distributions can directly provide estimators for $metric@K$; on the other side, since this question is closely related to the underlying mechanism of sampling for recommendation, tackling it can help better understand the power of sampling and help resolve the questions as to if and how we should use sampling for evaluating recommendation. 

Furthermore, since $metric@K$ is the linear function of $(P(R))_{R=1}^K$, the statistical properties of estimator $\widehat{P}(R)$ can be nicely preserved by $\widehat{metric}@K$~\cite{theorypoint}. 
In addition, this approach can be considered as metric-independent: We only need to estimate the empirical rank distribution $P(R)$ once; then we can utilize it for estimating all the top-$K$ evaluation metrics $metric@K$ (including for different $K$) based on Equation~\ref{eq:metric}. 

Finally, we note that we can utilize the $BV$ estimator to estimate $P(R)$ as follows: Let $\widehat{Recall}_{BV}(R)$ be the recall estimator from $BV$. Then we have 
\begin{small}
\begin{equation} \label{eq:BVR}
\begin{split}
   \widehat{P}(R) &=\widehat{Recall}_{BV}(R)-\widehat{Recall}_{BV}(R-1) \\
   &= (\tilde{P}(r))_{r=1}^n \Big((1.0-\gamma)A^TA+\gamma \text{diag}(\pmb{c})  \Big)^{-1}A^T\pmb{b}_R
\end{split}
\end{equation}
\end{small}

where $\widehat{Recall}_{BV}(R)$ is the $BV$ estimator for the $Recall@R$ metric, $(\tilde{P}(r))_{r=1}^n$ is the row vector of empirical rank distribution over the sampling data, and $\pmb{b}_R$ has the $R$-th element as $b_R$ (\cref{eq:BV_param}) and other elements as $0$. 
We consider this as our baseline for learning the empirical rank distribution.

\section{Learning Empirical Rank Distribution}\label{ch:estimators}
\begin{figure}
\centering
\begin{subfigure}[b]{0.5\textwidth}
   \includegraphics[width=1\linewidth]{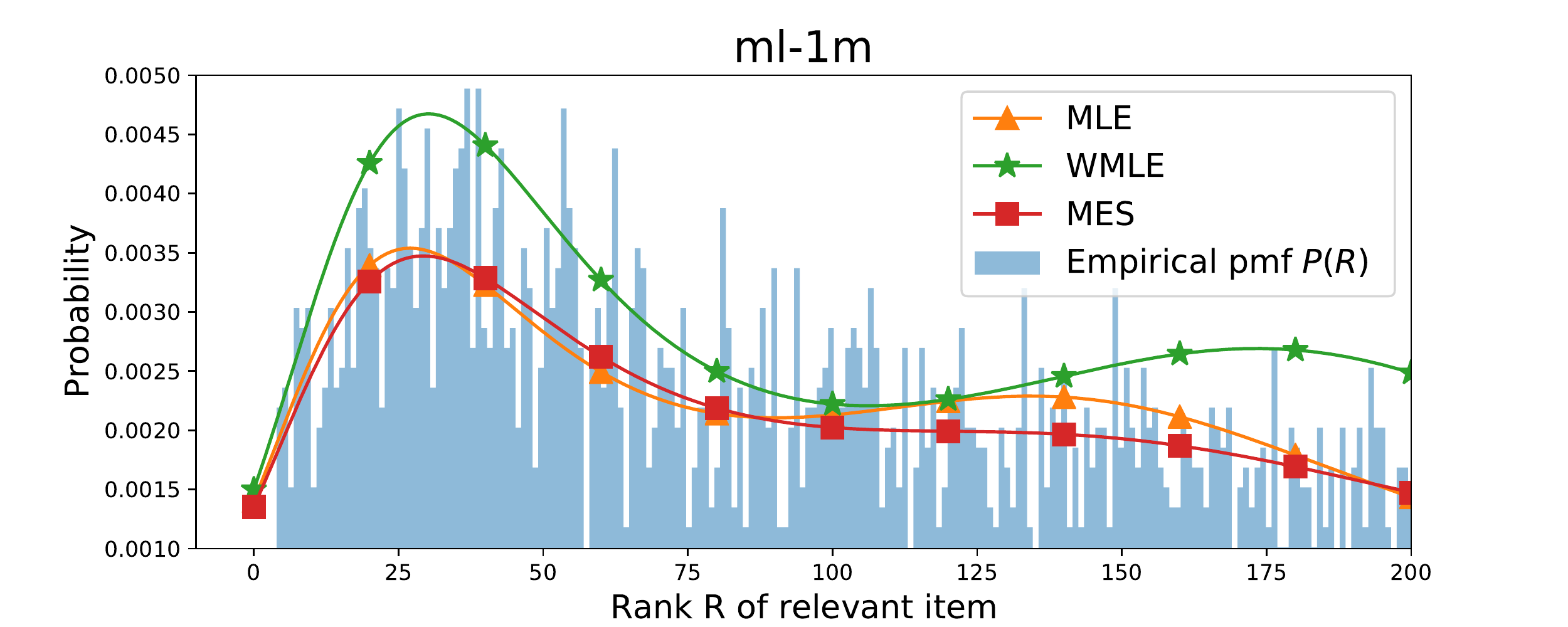}
   %\caption{}
   %\label{fig:Ng1} 
\end{subfigure}
% \begin{subfigure}[b]{0.55\textwidth}
%   \includegraphics[width=1\linewidth]{figure/distribution/pinterest-20}
%   %\caption{}
%   %\label{fig:Ng2}
% \end{subfigure}
\caption{Learning Empirical Rank Distribution $P(R)$}
\label{fig:P(R)}
\end{figure}

In this section, we will introduce a list of estimators  for the empirical rank distribution $(P(R))_{R=1}^N$ based on sampling ranked data: $\{r_u\}_{r=1}^M$. 
Figure~\ref{fig:P(R)} illustrates the different approaches of learning the empirical rank distribution $P(R)$, including the Maximal Likelihood Estimation (MLE), its weighted variants (WMLE), and the Maximal Entropy based approach (MES),  for $R \leq 200$ on movie-lens-1M dataset~\cite{ml-1m}.

%the sampling empirical rank distribution $(P(r))_{r=1}^n$, where $P(r)=\frac{1}{M} \sum_{u=1}^M {\bf 1}_{r_u=r}$.

% \subsection{Bayesian Nonparametric Estimator}\label{ch:BNE}
% ~\cite{Nonparametric book}

% \subsubsection{Rank Order Problem}
% This method is denoted as \textbf{BNE}.

\subsection{Sampling Rank Distribution: Mixtures of Binomial Distributions}

To simplify our discussion, let us consider the sampling with replacement scheme (the results can be extended to sampling without replacement). Now, assume an item $i$ is ranked $R$ in the entire set of items $I$. Then there are $R-1$ items whose rank is higher than item $i$ and $N-R$ items whose rank is lower than $i$. Under the (uniform) sampling (sampling with replacement), we have $\theta \coloneqq \frac{R-1}{N-1}$ probability to pick up an item with higher rank than $R$. 
Let $x$ be the number of irrelevant items ranked in front of the relevant one, and $x=r-1$. Thus, 
the rank $r-1$ under sampling follows a binomial distribution: $r-1 \sim B(n-1, \theta)$, and the conditional rank distribution $P(r|R)$ is
\begin{equation}
    \begin{split}
        &P(r|R)=Bin(r-1;n-1,\theta)= \binom{n-1}{r-1}\theta^{r-1}(1-\theta)^{n-r} 
    \end{split}
\end{equation}

Given this, an interesting observation is that the sampling ranked data $\{r_u\}_{r=1}^M$ can be directly modeled as a mixture of binomial distributions. 
Let $\pmb{\Theta}=(\theta_1\dots,\theta_R,\dots,\theta_N)^T$ where 
\begin{equation}
    \theta_R \coloneqq \frac{R-1}{N-1},\quad R = 1, \dots, N
\end{equation}
Let the empirical rank distribution $\pmb{P}=(P(R))_{R=1}^N$, then the sampling rank follows the distribution $P(r|\pmb{P})=$
\begin{small}
\begin{align}
&\sum_{R=1}^N P(r|R)P(R)=\sum_{R=1}^N Bin(r-1; n-1, \theta_R) P(R) \nonumber \\ 
=&\sum_{R=1}^N  P(R) \binom{n-1}{r-1} \big(\frac{R-1}{N-1}\big)^{r-1}\big(1-\frac{R-1}{N-1}\big)^{n-r} 
\end{align}
\end{small}
Thus, $P(R)$ can be considered as the parameters for the mixture of binomial distributions.

% For simplicity, we will define $x_u=r_u-1$, which is the number of irrelevant items ranked in front of the relevant one. We will consider $\{x_u\}_{u=1}^{M}$ and $\{r_u\}_{r=1}^M$ interchangeably as our observation. 

% Note, in this case,

%\subsubsection{Model Interpretation}

% \begin{equation}
%     p(\mathbf{z}|\pmb{\pi}) = \prod_{k=1}^{N}{{\pi^{z_k}_k}},\quad \pmb{z} \sim \text{Mult}(\pmb{\pi})
% \end{equation}

% where $\pmb{\Pi} = (\pi_1,\dots,\pi_k,\dots,\pi_N)^T$ and $\mathbf{z} = (z_1,\dots,z_k,\dots,z_N)^T$, $z_k\in\{0,1\}$, $\sum_k{z_k}=1$.

% \begin{equation}
%     p(x|\pmb{z}, \pmb{\theta}) = \prod_{k=1}^{N}{p(x|\theta_k)^{z_k}}
% \end{equation}
% \begin{equation}
% \begin{split}
% &p(x,\pmb{z}|\pmb{\theta})=\prod_{k=1}^{N}{\big[\pi_k p(x|\theta_k)\big]^{\pi_k}}\\
%     &p(x,z_k|\theta_k)=\pi_k p(x|\theta_k)
%     \end{split}
% \end{equation}
% \begin{equation}
% \begin{split}
%     &p(x|\theta_k) = \binom{n-1}{x}{\theta_k}^x(1-\theta_k)^{n-1-x}\\
%      &x|\theta_k \sim \text{Bin}(n-1, \theta_k)
%      \end{split}
% \end{equation}

\subsection{Maximum Likelihood Estimation}
The basic approach to learn the parameters of the mixture of binomial distributions ($MB$) given $\{r_u\}_{u=1}^M$ is based on maximal likelihood estimation (MLE). 
Let $\pmb{\Pi} = (\pi_1,\dots,\pi_R,\dots,\pi_N)^T$ be the parameters of the mixture of binomial distributions. Then we have $p(r_u|\pmb{\Pi})=\sum_{R=1}^N \pi_R p(r_u|\theta_R)$, where 
$p(r_u|\theta_R)=Bin(r_u-1; n-1, \theta_R)$. 

Then MLE aims to find the particular $\pmb{\Pi}$, which maximizes the log-likelihood: 
\begin{small}
\begin{equation}\label{eq:mb_mle}
    \log\mathcal{L} = \sum\limits_{u=1}^{M}{\log{p(r_u|\pmb{\Pi})}} = \sum\limits_{u=1}^{M}{ \log{\sum_{R=1}^N \pi_R p(r_u|\theta_R)}}
\end{equation}
\end{small}

%\frac{1}{M}$.

%   system cares more about the performance on the top of the ranking list, we introduce the weighted log-likelihood. Then our goal is trying to find $\pmb{\pi}$ by maximizing the weighted log-likelihood function:
% the empirical rank distribution $(P(R))_{R=1}^N$, we can utilize the 
% Since recommender system cares more about the performance on the top of the ranking list, we introduce the weighted log-likelihood. Then our goal is trying to find $\pmb{\pi}$ by maximizing the weighted log-likelihood function:

% \begin{small}
% \begin{equation}\label{eq:mb_mle}
% \begin{split}
%     \log\mathcal{L} &= \sum\limits_{u=1}^{M}{w_u\log{p(x_u|\pmb{\theta})}}\\
%     &=\sum\limites_{u=1}^{M}{w_u\log\sum\limits_{k=1}^{N}{p(x_u,{z}_{uk}}|{\theta}_k)}
%     \end{split}
% \end{equation}
% \end{small}

% \begin{equation}\label{eq:mb2}
%     M_k = \sum\limits_{u=1}^{M}{\gamma(z_{uk})}
% \end{equation}

By leveraging EM algorithm (see Appendix for details), we have:
\begin{equation}\label{eq:em_pi_new}
    \pi^{new}_R = \frac{1}{M} \sum\limits_{u=1}^M \frac{{\pi}^{old}_R p(r_u|\theta_R)}{\sum\limits_{j=1}^{N}{ \pi^{old}_j p(r_u|\theta_j)}}  
\end{equation} 

When \cref{eq:em_pi_new} converges, we obtain $\pmb{\Pi}^*$ and use it to estimate $\pmb{P}$, i.e., $\widehat{P}(R)=\pi^*_R$. 
Then, we can use $\widehat{P}(R)$ in \cref{eq:metric} to estimate the desired metric $metric@K$. 

\subsubsection{Speedup and Time Complexity}
To speedup the computation, we can further rewrite the updated formula \cref{eq:em_pi_new} as 
\begin{equation}\label{eq:em_pi_speedup}
    \pi^{new}_R = \sum\limits_{r=1}^n \tilde{P}(r) \frac{{\pi}^{old}_R p(r|\theta_R)}{\sum\limits_{j=1}^{N}{\pi^{old}_j p(r|\theta_j)}}
\end{equation} 
where $\tilde{P}(r)=\frac{1}{M} \sum_{u=1}^M {\bf 1}_{r_u=r}$ is the empirical rank distribution on the sampling data. 
Thus the time complexity improves to $O(kNn)$ (from $O(kNM)$ using \cref{eq:em_pi_new}) where $k$ is the iteration number. This is faster than the least squares solver for the $BV$ estimator (\cref{eq:leastsquare}) ~\cite{Krichene20@KDD20}, which is at least $O(n^2N)$.%DONG
Furthermore, we note $\widehat{P}(R)$ can be used for any $metric@K$ for the same algorithm, whereas $BV$ estimator has to be performed for each $metric@K$ separately.

\subsubsection{Weighted MLE}
If we are particularly interested in $\pi_R$ ($P(R)$) when $R$ is very small (such as $R<10$), then we can utilize the weighted MLE to provide more focus on those ranks. This is done by putting more weight on the sampling rank observation $r_u$ when $r_u$ is small. Specifically, the weighted MLE aims to find the $\pmb{\Pi}$, which maximizes the weighted log-likelihood: 

\begin{small}
\begin{equation}\label{eq:mb_mle}
    \log\mathcal{L} =  \sum\limits_{u=1}^{M}{w(r_u)\log{p(r_u|\pmb{\Pi})}} = \sum\limits_{u=1}^{M}{w(r_u) \log{\sum_{R=1}^N \pi_R p(r_u|\theta_R)}}
\end{equation}
\end{small}
where $w(r_u)$ is the weight for user $u$. Note that the typical MLE (without weight) is the special case of \cref{eq:mb_mle} ($w(r_u)=1$).

For weighted MLE, its updated formula is
\begin{equation}\label{eq:em_pi_speedup}
    \pi^{new}_R = \sum\limits_{r=1}^n  \frac{\tilde{P}(r) w(r)}{\sum_{r=1}^n \tilde{P}(r) w(r)}  \frac{{\pi}^{old}_R p(r|\theta_R)}{\sum\limits_{j=1}^{N}{\pi^{old}_j p(r|\theta_j)}}
\end{equation} 

For the weight $w_u$, we can utilize any decay function (as $r_u$ becomes bigger, than $w_u$ will reduce). We have experimented with various decay functions and found that the important/metric functions, such as $AP$ and $NDCG$,  $w_u=\mathcal{M}_{AP}(r_u/C)$ and  $w_u=\mathcal{M}_{NDCG}(r_u/C)$ ($C>1$ is a constant to help reduce the decade rate), obtain good and competitive results. We will provide their results in the experimental evaluation section. 

% \subsection{Mixtures of Binomial Model with Prior}
% Besides the mixture of binomial model, we could add a prior for $\pmb{\pi}$:
% \begin{equation}
%     p(\pmb{\pi}|\pmb{\alpha})=\frac{\Gamma(\alpha_0)}{\Gamma(\alpha_1)\cdots\Gamma(\alpha_N)}\prod_{k=1}^{N}{\pi^{\alpha_k-1}_k},\quad \pmb{\pi}\sim \text{Dir}(\pmb{\alpha})
% \end{equation}
% where $\pmb{\alpha}=(\alpha_1,\dots,\alpha_N)^T$, $\alpha_0=\sum_k{\alpha_k}$ is the parameter of Dirichlet distribution.
% The model is illustrated in \ref{fig:model_mb_prior}:
% \subsubsection{Maximum A Posteriori (MAP) Estimation}
% In this case, with the knowledge of prior distribution including its parameters, we are able to maximize the posterior distribution. 
% Still leveraging the EM algorithm (see Appendix for derivation), we have:
% \begin{equation}\label{eq:em_pi_new_prior}
%      \pi^{new}_k = \frac{ \sum\limits_{u=1}^{M}{w_u\cdot\gamma(z_{uk})} + \alpha_k - 1}{1+\alpha_0-N}
% \end{equation}
% where $\gamma(z_{uk})$ is the same as \cref{eq:em_gamma_zuk} and Repeat \cref{eq:em_gamma_zuk,eq:em_pi_new_prior} till convergence to obtain $\widehat{P}(R)$, then to estimate \cref{eq:empirical_metric}.

\subsection{Maximal Entropy with Minimal Distribution Bias}
Another commonly used approach for estimating a (discrete) probability distribution is based on the principal of maximal entropy~\cite{elementsinfo}. 
Assume a random variable $x$ takes values in $(x_1, x_2, \cdots, x_n)$ with pmf: $p(x_1), p(x_2), \cdots, p(x_n)$. Typically, given a list of (linear) constraints in the form of $\sum_{i=1}^n p(x_i) f_k(x_i) \geq F_k$ ($k=1, \cdots m$), together with the equality constraint ($\sum_{i=1}^n p(x_i)=1$), it aims to maximize its entropy: 
\begin{equation}
    H(p)=-\sum_{i=1}^{n} p(x_i) \log p(x_i)
\end{equation}

%Now, let us map our problem, estimating $P(R)$. 
In our problem, let the random variable $\mathcal{R}$ take on rank from $1$ to $N$. Assume its pmf is  $\pmb{\Pi}=(\pi_1,\dots,\pi_R,\dots,\pi_N)$, and the only immediate inequality constraint is $\pi_R \geq 0$ besides $\sum_{R=1}^N \pi_R=1$. 
Now, to further constrain $\pmb{\pi}$, we need to consider how they reflect and manifest on the observation data $\{r_u\}^M_{u=1}$. The natural solution is to simply utilize the (log) likelihood. However, combining them together leads to a rather complex non-convex optimization problem which will complicate the EM-solver. 

In this paper, we introduce a method (to constrain the maximal entropy) which utilizes the squared distance between the learned rank probability (based on $\pmb{\Pi}$) and the empirical rank probability in the sampling data

\begin{equation}\label{eq:entropy}
\begin{split}
    \mathcal{E}&= \frac{1}{M} \sum\limits_{R = 1}^{M}{\Big(p(r_u|\pmb{\Pi}) - \tilde{P}(r_u)  \Big)^2 } \\
    &=\sum\limits_{r = 1}^{n}{\tilde{P}(r)\Big(\sum\limits_{R = 1}^{N}{P(r|R)\pi_R} - \tilde{P}(r)  \Big)^2 }
    \end{split}
\end{equation}
Again, $\tilde{P}(r)$ is the empirical rank distribution in the sampling data. 
Note that $\mathcal{E}$ can be considered to be derived from the  log-likelihood of independent Gaussian distributions if we assume the error term $p(r_u|\pmb{\Pi}) - \tilde{P}(r_u)$ follows the Gaussian distribution.  

Given this, we seek to solve the following optimization problem: 
\begin{equation} \label{eq:MEE}
     \pmb{\Pi}=\arg\max_{\pmb{\Pi}} \eta\cdot H(\pmb{\pi})-\mathcal{E}  
\end{equation}
with constraints: 
\begin{equation}
    \pi_R \ge 0 \ (1 \leq R \leq N) \ \ \ \  \sum_R{\pi_R}=1
\end{equation}
Note that this objective can also be considered as adding an entropy regularizer for the log-likelihood. 

%\begin{lemma}
The objective function: $\eta\cdot H(\pmb{\pi})-\mathcal{E}$ is concave (or its negative is convex). 
%\end{lemma}
This can be easily observed as both negative of entropy and sum of squared errors are convex function. 

Given this, we can employ available convex optimization solvers~\cite{CVX} to identify the optimization solution. 
Thus, we have the estimator  $\widehat{P}(R)=\pi^*_R$, where $\Pi^*$ is the optimal solution for  \cref{eq:MEE}.

% \subsection{Minimal Cumulative Distribution Bias Estimator}
% In the paper \cite{}, there is mapping function $f(k)$, such that:
% \begin{equation}
%     CDF^{(r)}(k) = CDF^{(R)}(f(k))
% \end{equation}
% where $CDF^{(r)}$ and $CDF^{(R)}$ denote cumulative distribution function for $r$ and $R$ respectively.

% Recall that our goal is trying to estimate $P(R)$, further $\E{[\mathcal{M}(R)]}$.
% we note $\pmb{\pi} =[\pi_1,\dots,\pi_R,\dots,\pi_N]^T =[P(1),\dots,P(R),\dots,P(N)]^T$ is the vector of $P(R)$ distribution.
% \begin{equation}\label{eq:entropy}
% \begin{split}
%     \mathcal{L}&=\sum\limits_{r^\prime = 1}^{n}{P(r^\prime)\Big(  CDF^{(R)}(f(r^\prime))- \widetilde{CDF}^{(r)}(r^\prime) \Big)^2 }
%     \end{split}
% \end{equation}

% where $\widetilde{CDF}^{(r)}(r\prime)$ is the empirical cumulative distribution that can be obtained from observations $\{r_u\}_{u=1}^{M}$. 
% \begin{equation}
%     CDF^{(R)}(f(r^\prime)) = \sum\limits_{R=1}^{f(r^\prime)}{\pi_R}
% \end{equation}
% Then we could add entropy as a regularization term:
% \begin{equation}
%     \mathcal{L}^\prime = \mathcal{L} -\eta\cdot H(\pmb{\pi})=\mathcal{L}+\eta\cdot\pi_R\log \pi_R
% \end{equation}

% \begin{equation}
%     \pmb{\pi} = \arg\min_{\pmb{\pi}}{\mathcal{L}^\prime}
% \end{equation}
% with constraints:
% \begin{equation}
%     \pi_R\ge0,\sum_R{\pi_R}=1
% \end{equation}

\begin{scriptsize}
\begin{table*}[t]
\centering
\footnotesize
\begin{tabular}{|l|l|l|l|l|l|l|l|l|}
\hline
\multirow{2}{*}{Model}    & \multirow{2}{*}{Metric} & \multirow{2}{*}{Exact} & \multicolumn{6}{c|}{Estimators of $Metrics@10$}                                   \\ \cline{4-9} 
                          &                         &                       &CLS & BV 0.1        & BV 0.01       & MLE            & WMLE         & MES\\ \hline
\multirow{3}{*}{EASE}     & Recall                  & 34.77             & 54.43$\rpm$1.29     & 36.83$\rpm$2.09 & 36.18$\rpm$5.35 & 35.33$\rpm$7.17 & 36.30$\rpm$7.41 & 35.22$\rpm$7.36  \\ \cline{2-9} 
                          & NDCG                    & 16.16      & 25.44$\rpm$0.60            & 16.81$\rpm$1.04 & 16.38$\rpm$2.88 & 16.03$\rpm$3.95 & 16.46$\rpm$4.07 & 16.03$\rpm$4.12  \\ \cline{2-9} 
                          & AP                      & 10.63      & 16.81$\rpm$0.40            & 10.88$\rpm$0.72 & 10.53$\rpm$2.13 & 10.32$\rpm$2.97 & 10.59$\rpm$3.06 & 10.35$\rpm$3.13  \\ \hline
\multirow{3}{*}{MultiVAE} & Recall                  & 18.38        & 45.23$\rpm$1.42          & 26.27$\rpm$2.46 & 21.66$\rpm$6.11 & 20.79$\rpm$6.78 & 21.23$\rpm$6.96 & 20.58$\rpm$6.97  \\ \cline{2-9} 
                          & NDCG                    & 7.08       & 21.13$\rpm$0.66            & 11.80$\rpm$1.22 & 9.38$\rpm$3.26  & 9.17$\rpm$3.44  & 9.35$\rpm$3.53  & 9.10$\rpm$3.58   \\ \cline{2-9} 
                          & AP                      & 3.81        & 13.97$\rpm$0.44           & 7.53$\rpm$0.85  & 5.77$\rpm$2.39  & 5.74$\rpm$2.44  & 5.86$\rpm$2.50  & 5.72$\rpm$2.56   \\ \hline
\multirow{3}{*}{NeuMF}    & Recall                  & 30.96         & 49.51$\rpm$1.31        & 32.12$\rpm$2.17 & 31.30$\rpm$5.63 & 31.06$\rpm$7.15 & 31.82$\rpm$7.37 & 30.63$\rpm$7.30  \\ \cline{2-9} 
                          & NDCG                    & 13.43      & 23.14$\rpm$0.61           & 14.62$\rpm$1.08 & 14.15$\rpm$3.03 & 14.16$\rpm$3.94 & 14.49$\rpm$4.05 & 13.99$\rpm$4.06  \\ \cline{2-9} 
                          & AP                      & 8.26       & 15.29$\rpm$0.40            & 9.44$\rpm$0.75  & 9.08$\rpm$2.24  & 9.15$\rpm$2.96  & 9.37$\rpm$3.04  & 9.06$\rpm$3.07   \\ \hline
\multirow{3}{*}{itemKNN}  & Recall                  & 42.72         & 46.46$\rpm$1.28         & 34.26$\rpm$2.00 & 38.29$\rpm$5.09 & 40.02$\rpm$7.22 & 41.71$\rpm$7.56 & 39.20$\rpm$6.43  \\ \cline{2-9} 
                          & NDCG                    & 20.54      & 21.71$\rpm$0.60          & 15.80$\rpm$0.99 & 17.81$\rpm$2.74 & 18.96$\rpm$4.24 & 19.75$\rpm$4.43 & 18.53$\rpm$3.73  \\ \cline{2-9} 
                          & AP                      & 13.89      & 14.35$\rpm$0.40           & 10.32$\rpm$0.69 & 11.73$\rpm$2.02 & 12.68$\rpm$3.32 & 13.21$\rpm$3.47 & 12.38$\rpm$2.90  \\ \hline
\multirow{3}{*}{ALS}      & Recall                  & 24.17        & 48.07$\rpm$1.20          & 29.62$\rpm$2.04 & 26.17$\rpm$5.64 & 25.16$\rpm$6.49 & 25.91$\rpm$6.72 & 24.94$\rpm$7.08  \\ \cline{2-9} 
                          & NDCG                    & 9.49       & 22.46$\rpm$0.56           & 13.39$\rpm$1.02 & 11.54$\rpm$3.08 & 11.18$\rpm$3.40 & 11.51$\rpm$3.52 & 11.14$\rpm$3.76  \\ \cline{2-9} 
                          & AP                      & 5.21        & 14.84$\rpm$0.37          & 8.59$\rpm$0.72  & 7.23$\rpm$2.30  & 7.06$\rpm$2.47  & 7.27$\rpm$2.55  & 7.08$\rpm$2.76   \\ \hline
\end{tabular}
\caption{Dataset: ml-1m with sample size =$99$.}
\label{tab:ml-1m_base}
\end{table*}
\end{scriptsize}

\begin{scriptsize}
\begin{center}
\begin{table*}[t]
\centering
\footnotesize
\begin{tabular}{|l|l|l|l|l|l|l|l|l|}
\hline
\multirow{2}{*}{Model}    & \multirow{2}{*}{Metric} & \multirow{2}{*}{Exact} & \multicolumn{6}{c|}{Estimators of $Metric@10$}                                     \\ \cline{4-9} 
                          &                         &             & CLS           & BV 0.1        & BV 0.01       & MLE             & WMLE          & MES \\ \hline
\multirow{3}{*}{EASE}     & Recall                  & 87.91        & 52.56$\rpm$0.52          & 49.62$\rpm$1.06 & 65.99$\rpm$2.98 & 83.19$\rpm$10.14 & 84.13$\rpm$10.26 & 83.72$\rpm$11.83 \\ \cline{2-9} 
                          & NDCG                    & 43.63        & 24.04$\rpm$0.24              & 22.70$\rpm$0.49 & 30.28$\rpm$1.39 & 38.55$\rpm$4.97  & 38.99$\rpm$5.03  & 38.83$\rpm$5.81  \\ \cline{2-9} 
                          & AP                      & 30.48         & 15.58$\rpm$0.15             & 14.73$\rpm$0.32 & 19.70$\rpm$0.92 & 25.29$\rpm$3.42  & 25.58$\rpm$3.46  & 25.49$\rpm$4.01  \\ \hline
\multirow{3}{*}{MultiVAE} & Recall                  & 48.82           & 55.62$\rpm$0.54           & 52.84$\rpm$1.10 & 70.09$\rpm$2.96 & 86.45$\rpm$9.89  & 86.98$\rpm$9.95  & 87.02$\rpm$10.82 \\ \cline{2-9} 
                          & NDCG                    & 17.12        & 25.43$\rpm$0.25            & 24.17$\rpm$0.51 & 32.16$\rpm$1.38 & 39.98$\rpm$4.82  & 40.22$\rpm$4.85  & 40.27$\rpm$5.29  \\ \cline{2-9} 
                          & AP                      & 8.14         & 16.49$\rpm$0.16             & 15.68$\rpm$0.33 & 20.92$\rpm$0.91 & 26.18$\rpm$3.30  & 26.34$\rpm$3.32  & 26.39$\rpm$3.63  \\ \hline
\multirow{3}{*}{NeuMF}    & Recall                  & 62.87            & 45.83$\rpm$0.56         & 41.51$\rpm$1.12 & 53.39$\rpm$3.16 & 63.68$\rpm$9.42  & 64.24$\rpm$9.52  & 64.39$\rpm$11.44 \\ \cline{2-9} 
                          & NDCG                    & 31.05        & 20.96$\rpm$0.26              & 18.98$\rpm$0.52 & 24.48$\rpm$1.48 & 29.41$\rpm$4.58  & 29.67$\rpm$4.62  & 29.77$\rpm$5.59  \\ \cline{2-9} 
                          & AP                      & 21.66        & 13.59$\rpm$0.17            & 12.30$\rpm$0.34 & 15.91$\rpm$0.98 & 19.23$\rpm$3.13  & 19.41$\rpm$3.16  & 19.50$\rpm$3.83  \\ \hline
\multirow{3}{*}{itemKNN}  & Recall                  & 68.46              & 52.88$\rpm$0.52        & 50.42$\rpm$1.03 & 68.04$\rpm$3.08 & 88.43$\rpm$11.35 & 89.36$\rpm$11.48 & 89.19$\rpm$13.13 \\ \cline{2-9} 
                          & NDCG                    & 28.71        & 24.18$\rpm$0.24             & 23.07$\rpm$0.48 & 31.23$\rpm$1.44 & 41.06$\rpm$5.59  & 41.49$\rpm$5.66  & 41.46$\rpm$6.49  \\ \cline{2-9} 
                          & AP                      & 17.12        & 15.68$\rpm$0.15           & 14.97$\rpm$0.31 & 20.32$\rpm$0.95 & 26.98$\rpm$3.86  & 27.27$\rpm$3.91  & 27.28$\rpm$4.49  \\ \hline
\multirow{3}{*}{ALS}      & Recall                  & 58.55            & 31.39$\rpm$0.48          & 26.90$\rpm$0.93 & 34.43$\rpm$2.62 & 42.21$\rpm$7.16  & 43.05$\rpm$7.31  & 43.09$\rpm$8.91  \\ \cline{2-9} 
                          & NDCG                    & 30.30         & 14.35$\rpm$0.22             & 12.29$\rpm$0.43 & 15.78$\rpm$1.22 & 19.55$\rpm$3.50  & 19.94$\rpm$3.57  & 20.00$\rpm$4.39  \\ \cline{2-9} 
                          & AP                      & 21.77         & 9.31$\rpm$0.14            & 7.97$\rpm$0.28  & 10.26$\rpm$0.81 & 12.82$\rpm$2.40  & 13.07$\rpm$2.45  & 13.14$\rpm$3.02  \\ \hline
\end{tabular}
\caption{Dataset: citeulike with sample size =$99$.}
\label{tab:citeulike_base}
\end{table*}
\end{center}
\end{scriptsize}
\section{Experiments}

\begin{figure*}
    \centering
    \includegraphics[width=\textwidth]{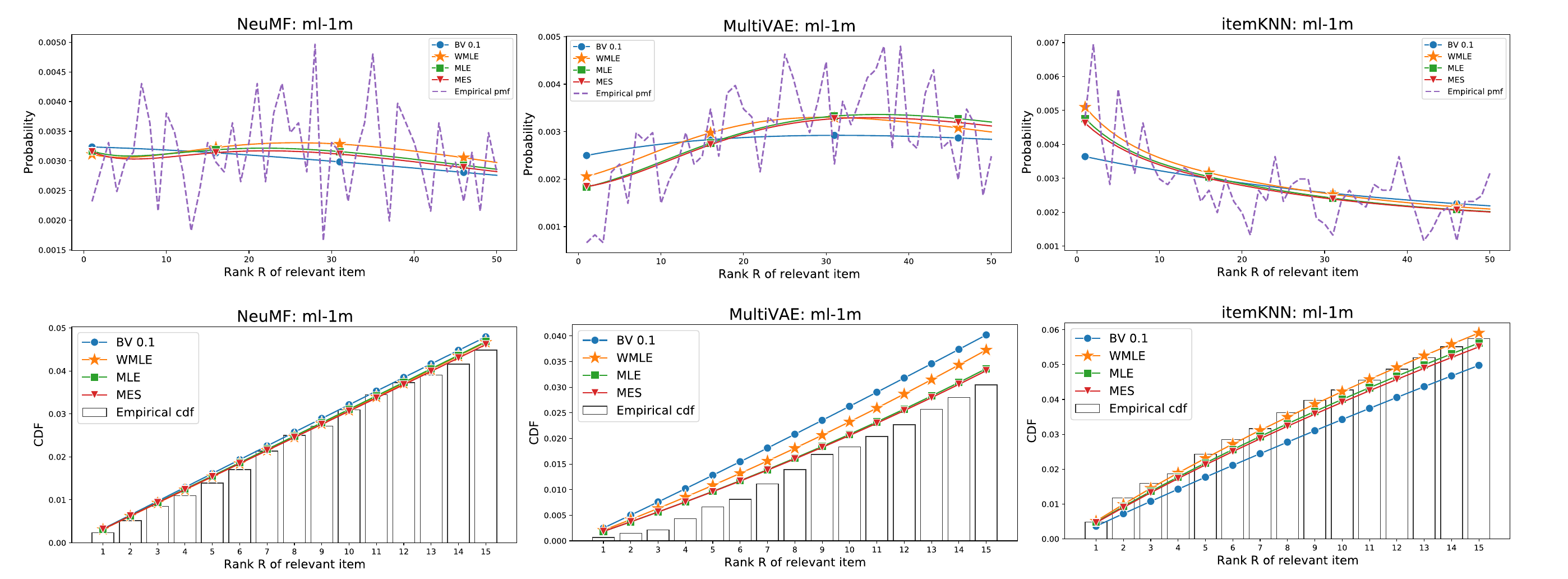}
    \vspace*{-3.0ex}
    \caption{Accuracy of Learned Empirical Rank Distribution}
    \label{fig:LERD}
    \vspace*{-3.0ex}
\end{figure*}

In this section, we report the experimental evaluation on  estimating the top-$K$ metrics based on sampling, as well as the learning of empirical rank distribution $P(R)$. 
Specifically, we aim to answer the following questions: 

% \begin{small}
% \begin{table}
%   \caption{Dataset Statistics}
%   \label{tab:datasets}
%   \begin{tabular}{lcccc}
%     \toprule
%     \textbf{Dataset}&
%     \textbf{Interactions}&
%     \textbf{Users}&
%     \textbf{Items}&\textbf{Sparsity}\\
%     \midrule
%     ml-1m  &1,000,209 & 6,040 &3,706&95.53$\%$\\
%     pinterest-20& 1,463,581 & 55,187&9,916&99.73$\%$\\
%     citeulike&204,986 &5,551&16,980&99.78$\%$ \\
%     yelp& 696,865&25,677 &25,815&99.89$\%$
%     \bottomrule
%   \end{tabular}
% \end{table}
% \end{small}

\noindent{(Question 1)} How do the new estimators based on the learned empirical distribution perform against the $CLS$ and $BV$ approach proposed in ~\cite{Krichene20@KDD20} on estimating the top-$K$ metrics based on sampling?

\noindent{(Question 2)} How do these approaches perform when helping predict the winners (from the global metrics) among a list of competitive recommendation algorithms using sampling?

\noindent{(Question 3)} How accurately can the proposed approaches learn the empirical rank distribution? 

% $(which potentially provides the lower bound for a specific algorithm on a given dataset, based on Beta distribution)

% In this section, we only focus on Question 1 and Question 4. The discussion of Question 2 and 3, together with their experimental results and the experimental setup will be put into the Appendix~\ref{ap:ex}. 

\subsection{Experimental Setup}
We use four of the most commonly used datasets for recommendation studies in our study, whose  characteristics are in the Appendix.
%in Table~\ref{tab:datasets}. 
For the different recommendation algorithms, we use some of the most well-known and the state-of-the-art algorithms, including three non-deep-learning options: itemKNN~\cite{DeshpandeK@itemKNN};
ALS~\cite{hu2008collaborative}; and EASE~\cite{Steck_2019}; 
and two deep learning options:   
NeuMF~\cite{he2017neural} and 
MultiVAE~\cite{liang2018variational}.
We use three (likely the most) commonly used top-K evaluation metrics: $Recall$, $NDCG$ and $AP$ (Average Precision). 
Due to the space limitation, we only report representative results here, and additional experimental results can be found in the Appendix. 

\subsection{Estimation Accuracy of Metric@$K$}
Table~\ref{tab:ml-1m_base} and~\ref{tab:citeulike_base} show the average and the standard deviation of the aforementioned estimators for $Recall@10$, $NDCG@10$, and $AP@10$, which repeats $100$ each with sample size $99$ ($n=100$). 
The estimators include $CLS$, $BV$ (with the tradeoff parameters $\gamma=0.1$ and $\gamma=0.01$), $MLE$ (Maximal Likelihood Estimation), $WMLE$ (Weighted Maximal Likelihood Estimation where the weighted function is $M_{NDCG}$ with $C=10$), $MES$ (Maximal Entropy with Squared distribution distance, where $\eta=0.001$). The $Exact$ column corresponds to the target metrics which use all items in $I$ for ranking.  

In Table~\ref{tab:ml-1m_base} on the $ml-1m$ dataset, we observe that $MLE$ and $MES$ are among the most, or the second-most accurate estimators (using the bias which measures the difference between the Exact Metrics and the Average of the Estimated Metrics). 
In Table~\ref{tab:citeulike_base}, $WMLE$ performs the best, with $7$ most (or second-most) accurate estimations, whereas $MES$, $MB$, and $BV$ estimators are all comparable, with each having some better estimates. In both tables,  $CLS$ estimator has the worst performance. 
%{\color{red} We also notice, in this dataset $CLS$ estimator gets better performance but still underperforms  others.}
In addition, we also notice that the new estimators tend to have higher variance than the $BV$ estimators, which explicitly control the variance of each individual ${\mathcal M}(r)$ estimate. 
In the future, we plan to utilize methods such as bootstrapping to help reduce the variance of these estimators based on the empirical rank distribution.

\subsection{Predicting Winners by Metric@$K$}
Table~\ref{tab:winner_ml} shows, among the $100$ sample runs, the number of correct winners predicted by $CLS$, $BV$, $MLE$, $WLE$ and $MES$ estimators based on $Recall@K$, $NDCG@K$ and $AP@K$ for $K=1$, $5$, $10$ and $20$, on the $ml-1m$ dataset. We observe that $WMLE$ has the best prediction accuracy in picking up the winners, while $MLE$ and $MES$  are comparable and slightly better than $BV0.01$.

\subsection{Learning Empirical Rank Distributions}
Figure~\ref{fig:LERD} illustrates the accuracy of learned empirical Rank Distributions against the exact $P(R)$ on the $ml-1m$ dataset for three recommendation methods: $NeuMF$, $MultiVAE$, and $itemKNN$, respectively. The {\em empirical pmf} refers to $P(R)$, and the estimation methods include $BV$ (with parameter $0.1$), $MLE$, $WMLE$, and $MES$. The three figures on the top show the (learned) probability mass function, whereas the bottom shows the corresponding CDF (or Recall) at the top-$K$. 
These estimation curves are the average of estimates of $100$ sample runs. 
We can see that $BV$ either over- or under- estimates the empirical CDF ($Recall$ curve). $MLE$ and $MES$ are quite comparable where $WMLE$ has a higher average estimate than both of them. This is understandable as we add more weight to the sampled rank with smaller values, which leads to a higher concentration of probability mass for the smaller rank. We also notice that all these estimators are not very accurate on the individual rank probability $P(R)$ (the top figures). But their aggregated results (CDF; the bottom figures) are quite accurate. This helps explain why we can estimate $metric@K$, which is also an aggregate. In the Appendix, we show that as the sample size increases, the estimate accuracy will also increase accordingly. 

% % Please add the following required packages to your document preamble:
% % \usepackage{multirow}
% \begin{scriptsize}
% \begin{table}[]
% \begin{tabular}{|c|l|l|l|l|l|l|}
% \hline
% K                   & Metric & BV 0.1 & BV 0.01 & MLE & WMLE & MES \\ \hline
% \multirow{3}{*}{1}  & Recall & 41    & 59     & 59 & 59    & 57             \\ \cline{2-7} 
%                     & NDCG   & 41    & 59     & 59 & 59    & 57             \\ \cline{2-7} 
%                     & AP     & 41    & 59     & 59 & 59    & 57             \\ \hline
% \multirow{3}{*}{5}  & Recall & 34    & 57     & 59 & 61    & 59             \\ \cline{2-7} 
%                     & NDCG   & 34    & 57     & 59 & 61    & 59             \\ \cline{2-7} 
%                     & AP     & 35    & 58     & 59 & 61    & 59             \\ \hline
% \multirow{3}{*}{10} & Recall & 21    & 54     & 58 & 59    & 58             \\ \cline{2-7} 
%                     & NDCG   & 24    & 56     & 60 & 61    & 59             \\ \cline{2-7} 
%                     & AP     & 27    & 56     & 61 & 61    & 59             \\ \hline
% \end{tabular}
% \caption{The number of successes at predicting a winner on the ml-1m dataset with 100 repeats.}
% \label{tab:winner_ml}
% \vspace{-3.0ex}
% \end{table}
% \end{scriptsize}

% \usepackage{multirow}
\begin{tiny}
\begin{table}[]
\footnotesize
\begin{tabular}{|c|c|c|c|c|c|c|c|}
\hline
K                   & Metric & CLS& 0.1 & 0.01 & MLE & WMLE & MES \\ \hline
\multirow{3}{*}{1}  & Recall & 0  & 41    & 59     & 59 & 59    & 57             \\ \cline{2-8} 
                    & NDCG   & 0  & 41    & 59     & 59 & 59    & 57             \\ \cline{2-8} 
                    & AP     & 0  & 41    & 59     & 59 & 59    & 57             \\ \hline
\multirow{3}{*}{5}  & Recall & 0  & 34    & 57     & 59 & 61    & 59             \\ \cline{2-8} 
                    & NDCG   & 0  & 34    & 57     & 59 & 61    & 59             \\ \cline{2-8} 
                    & AP     & 0  & 35    & 58     & 59 & 61    & 59             \\ \hline
\multirow{3}{*}{10} & Recall & 0  & 21    & 54     & 58 & 59    & 58             \\ \cline{2-8} 
                    & NDCG   & 0  & 24    & 56     & 60 & 61    & 59             \\ \cline{2-8} 
                    & AP     & 0  & 27    & 56     & 61 & 61    & 59             \\ \hline
\multirow{3}{*}{20} & Recall & 100  & 98    & 59     & 49 & 45    & 53             \\ \cline{2-8} 
                    & NDCG   & 0  & 4     & 51     & 54 & 58    & 53             \\ \cline{2-8} 
                    & AP    & 0   & 23    & 53     & 60 & 61    & 58             \\ \hline
\end{tabular}
\caption{The number of successes at predicting a winner on the ml-1m dataset with 100 repeats. 0.1 and 0.01 represent the estimator BV with $\gamma = 0.1$ and $0.01$ correspondingly.}
\label{tab:winner_ml}
\vspace{-3.0ex}
\end{table}
\end{tiny}

\section{Conclusion}
In this paper, we study a new approach to estimate the top-$K$ evaluation metrics based on learning the empirical  rank distribution from sampling, which is, by itself, a new and interesting research problem. We present two approaches based on Maximal Likelihood Estimation and Maximal Entropy principals. Our experimental results show the advantage of using the new approaches to estimate the top-$K$ metrics. In our future work, we plan to investigate the open questions on how many samples we should use for recovering the empirical rank distribution and top-$K$ metrics. 

\newpage
\bibliography{ref.bib}
\newpage
\hfill
%\newpage
\appendix
\section{Dataset Statistics}
In table ~\ref{tab:datasets}, we describe the information of the dataset we used.
\begin{scriptsize}
\begin{table}[h]
  \begin{tabular}{lcccc}
    \toprule
    \textbf{Dataset}&
    \textbf{Interactions}&
    \textbf{Users}&
    \textbf{Items}&\textbf{Sparsity}\\
    \midrule
    ml-1m  &1,000,209 & 6,040 &3,706&95.53$\%$\\
    pinterest-20& 1,463,581 & 55,187&9,916&99.73$\%$\\
    citeulike&204,986 &5,551&16,980&99.78$\%$ \\
    yelp& 696,865&25,677 &25,815&99.89$\%$\\
    \bottomrule
  \end{tabular}
  \caption{Dataset Statistics}
  \label{tab:datasets}
\end{table}
\end{scriptsize}

\section{EM Algorithm for Mixtures of Binomial Model}
In this section, we give the details of the EM algorithm for $MLE$ estimator.
Recall \cref{eq:mb_mle}
Weighted log-likelihood function is:
\begin{equation*}
    \log{\mathcal{L}} = \sum\limits_{u=1}^{M}{w_u \log{\sum\limits_{k=1}^{N}{p(x_u,z_{uk}|\theta_k)}}}
\end{equation*}

\subsubsection{E-step}

\begin{equation*}
\mathcal{Q}(\pmb{\pi},\pmb{\pi}^{old})=\sum\limits_{u=1}^{M}{w_u \sum\limits_{k=1}^{N}{\gamma(z_{uk}) \log{p(x_u,z_{uk}|\theta_k)}}}
\end{equation*}

where
\begin{equation*}
    \gamma(z_{uk})=p(z_{uk}|x_u,\pmb{\pi}^{old})=\frac{{\pi}^{old}_k p(x_u|\theta_k)}{\sum\limits_{j=1}^{N}{\pi^{old}_j p(x_u|\theta_j)}}
\end{equation*}

\subsubsection{M-step}
\begin{equation}\label{eq:Q_prime}
   \mathcal{Q}^\prime(\pmb{\pi},\pmb{\pi}^{old})=\mathcal{Q}(\pmb{\pi},\pmb{\pi}^{old}) + \lambda (1-\sum\limits_{k=1}^{N}{\pi_k})
\end{equation}

\begin{equation*}
    \frac{\partial \mathcal{Q}^\prime(\pmb{\pi},\pmb{\pi}^{old})}{\partial \pi_k}=\sum\limits_{u=1}^{M}{ \frac{w_u\gamma(z_{uk}) }{\pi_k} -\lambda}=0
\end{equation*}

\begin{equation*}
    \lambda\pi_k=\sum\limits_{u=1}^M{w_u\cdot\gamma(z_{uk})}
\end{equation*}

% \begin{equation*}
%     \lambda=\sum\limits_{k=1}^{N}\sum\limits_{u=1}^M{w_u\cdot\gamma(z_{uk})}=1
% \end{equation*}

% \begin{equation*}
%     \pi^{new}_k=M_k
% \end{equation*}

\begin{equation}
    \lambda=\sum\limits_{k=1}^{N}\sum\limits_{u=1}^M{w_u\cdot\gamma(z_{uk})}=\sum\limits_{u=1}^M{w_u}
\end{equation}

\begin{equation*}
    \pi^{new}_k=\frac{\sum\limits_{u=1}^M{w_u\cdot\gamma(z_{uk})}}{\sum\limits_{u=1}^M{w_u} }
\end{equation*}

\begin{figure*}
%\vspace*{-5in}
\centering
\begin{subfigure}{0.5\textwidth}
  \centering
  \includegraphics[width=1.0\linewidth,scale = 1.0]{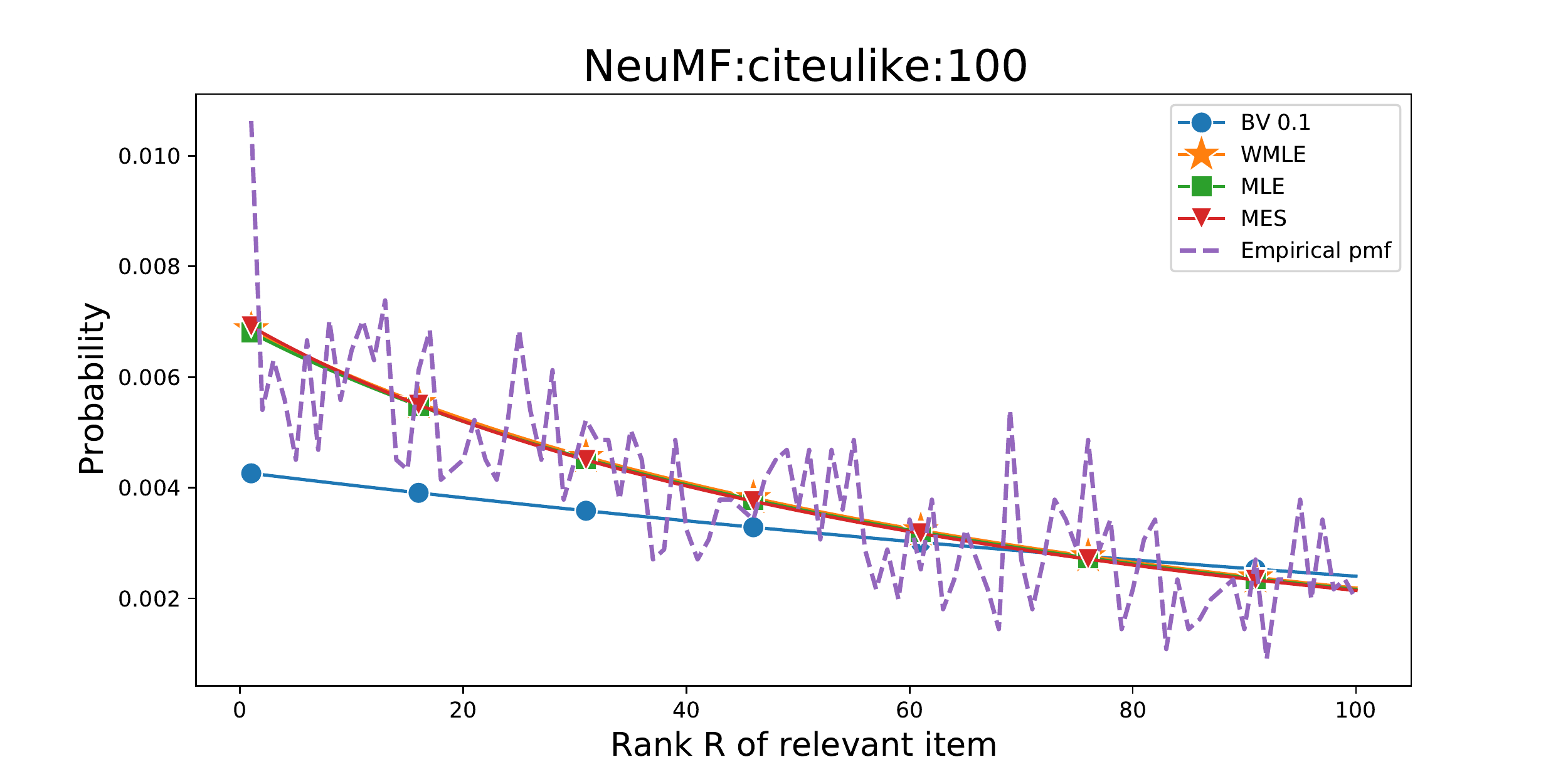}
  \caption{}
  \label{2a}
\end{subfigure}%
\begin{subfigure}{.5\textwidth}
  \centering
  \includegraphics[width=1.0\linewidth,scale = 1.0]{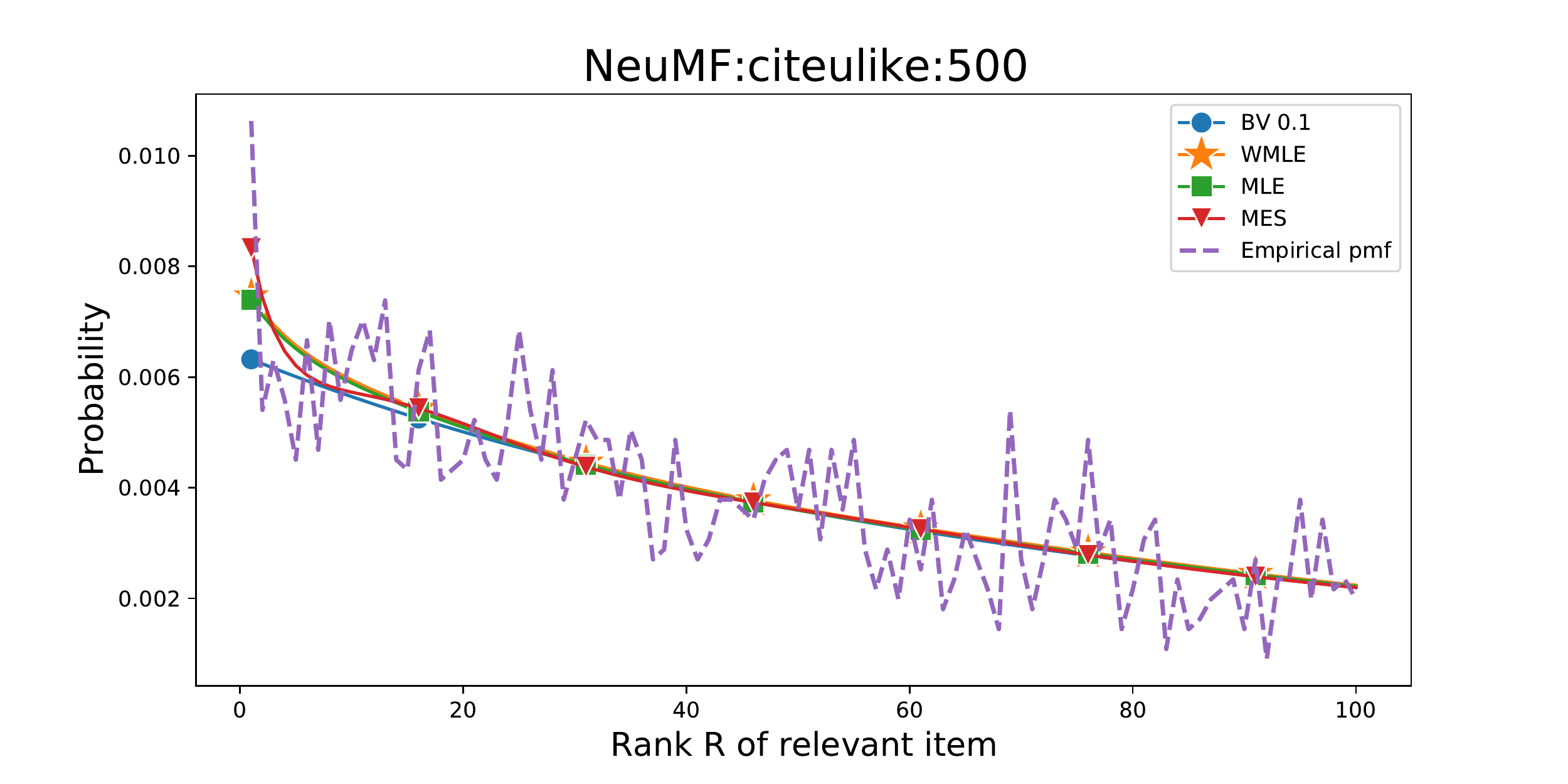}
  \caption{}
  \label{2b}
\end{subfigure}
\begin{subfigure}{0.5\textwidth}
  \centering
  \includegraphics[width=1.0\linewidth,scale = 1.0]{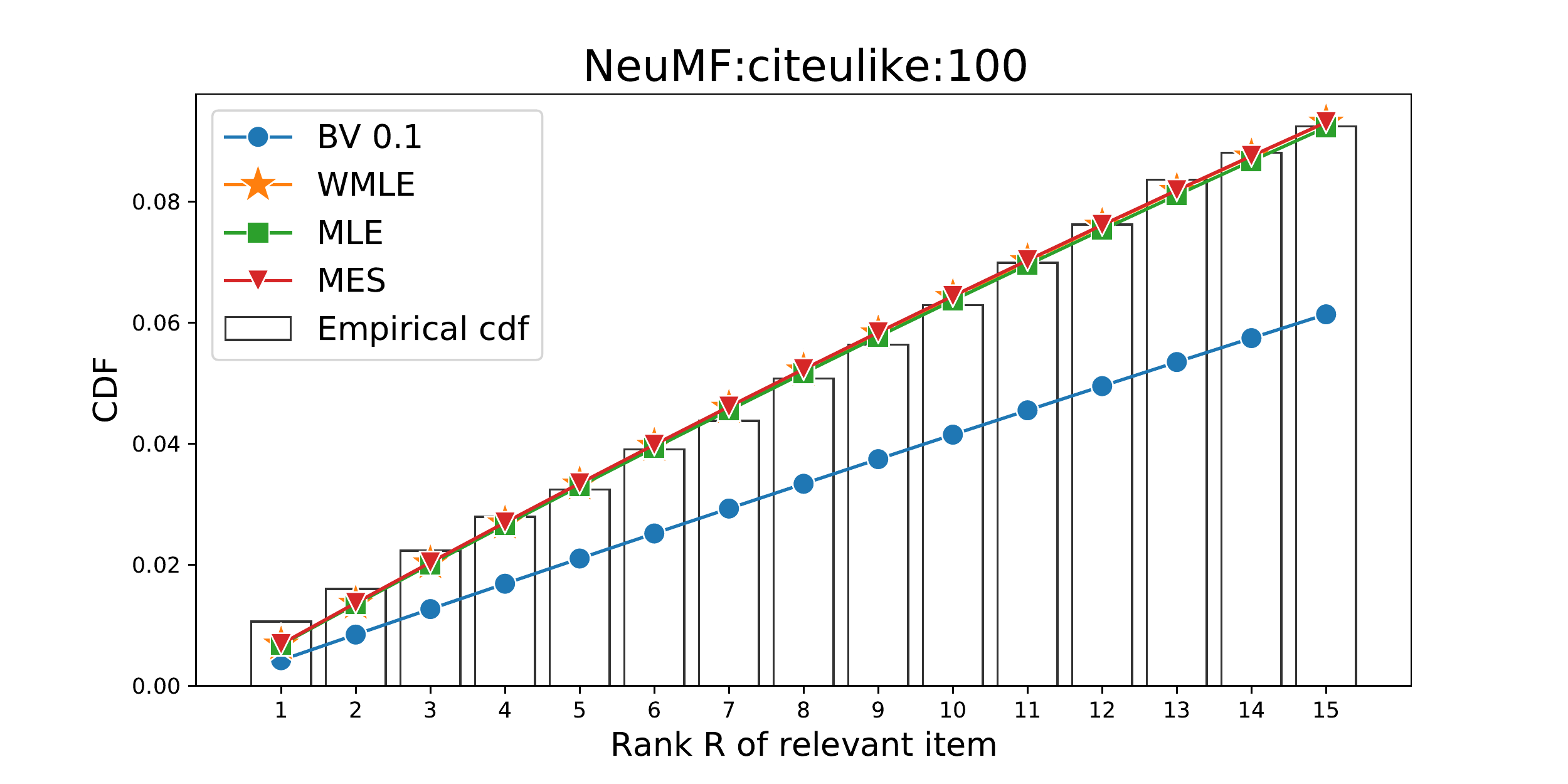}
  \caption{}
  \label{2c}
\end{subfigure}%
\begin{subfigure}{.5\textwidth}
  \centering
  \includegraphics[width=1.0\linewidth,scale = 1.0]{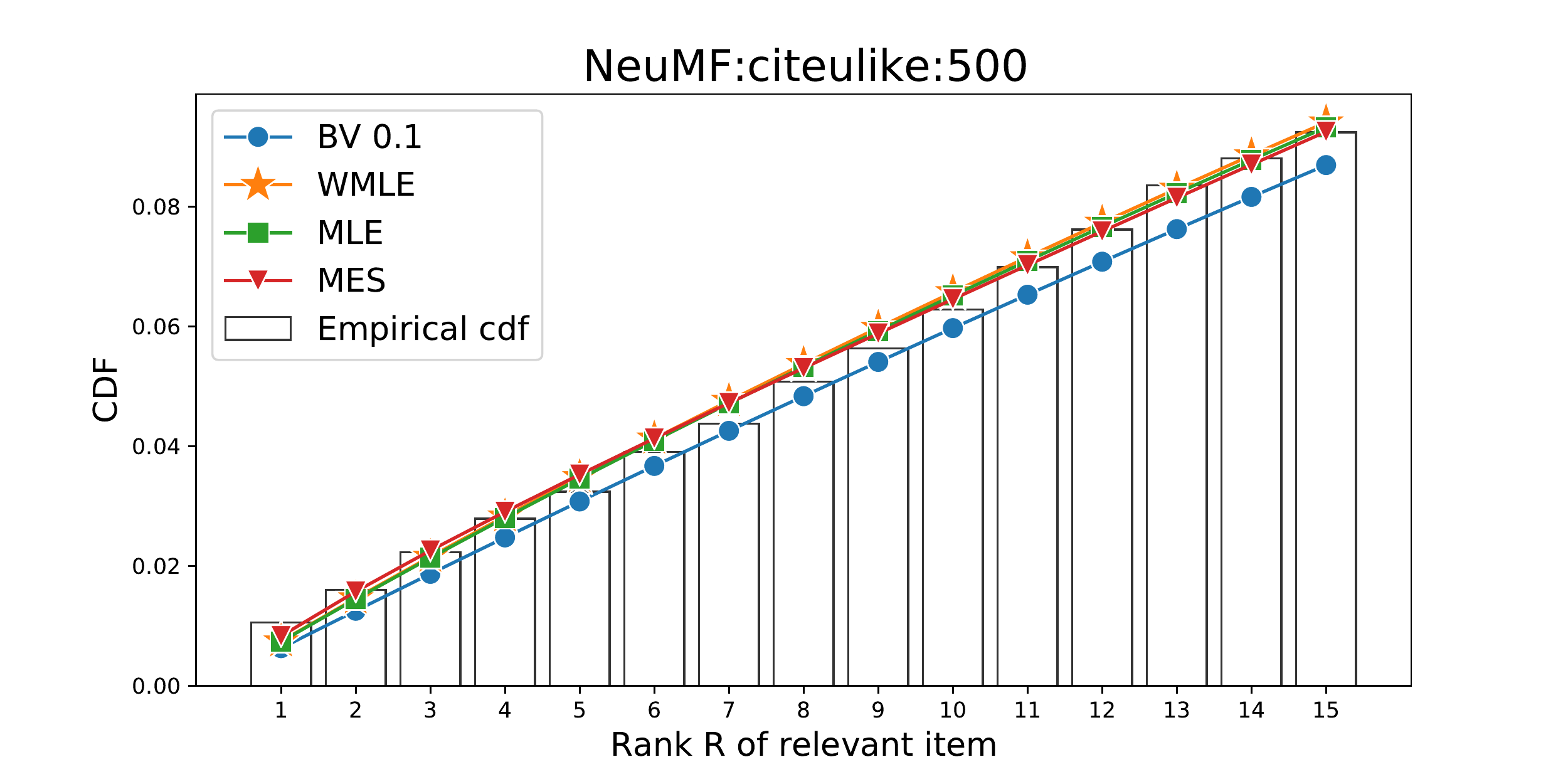}
  \caption{}
  \label{2d}
\end{subfigure}
\caption{Accuracy of Learned Empirical Rank Distribution with Different Sample Size. The experiments are carryied on dataset $citeulike$ with recommender $NeuMF$. $100$ and $500$ indicate the sample size is $99$ and $499$ respectively. As the sample size increases, the estimation curves tend to get closer to the empirical rank curve, especially for the $BV 0.1$ estimator.}
\label{fig:citeulike_100_500}
\end{figure*}

\section{Sample Size Impact Accuracy}
Figure~\ref{fig:citeulike_100_500} illustrates the accuracy of learned empirical Rank Distributions against the exact $P(R)$ on the $citeulike$ dataset for recommender $NeuMF$.
Figure ~\ref{2a} and ~\ref{2b} show the probability mass function with sample size 99 (n = 100) and 499 (n = 500) respectively. Similarly, Figure ~\ref{2c} and ~\ref{2d} display the corresponding CDF at the top-$K$. These estimation curves are the average of estimates of $100$ sample runs. 
From the figures, we can see that, as the sample size increases, the estimation curves get close to the empirical distribution. That's to say, as the sample size increases, the estimate accuracy will also increase accordingly.

\section{Estimation Accuracy of Metric@$K$}

Table~\ref{tab:pinterest_base} and ~\ref{tab:yelp_base} show the average and the standard deviation of the aforementioned estimators for $Recall@10$, $NDCG@10$, and $AP@10$, which runs $100$ samples each with sample size $99$ ($n=100$). 

In Table~\ref{tab:pinterest_base} on the $pinterest-20$ dataset and table~\ref{tab:yelp_base} on the $yelp$ dataset, we observe that $MLE$ and $MES$ are pretty much accurate than $BV$ estimator.  

Overall, $MLE$ and $MES$ can produce more accurate estimation of metrics than $BV$ with higher variance.

% \usepackage{multirow}
\begin{small}
\begin{table*}[t]
\begin{tabular}{|c|c|c|c|c|c|c|c|}
\hline
\multirow{2}{*}{Model}    & \multirow{2}{*}{Metric} & \multirow{2}{*}{Exact} & \multicolumn{5}{c|}{Estimators of $Metric@10$}                                   \\ \cline{4-8} 
                          &                         &                        & BV 0.1        & BV 0.01       & MLE           & WMLE        & MES \\ \hline
\multirow{3}{*}{EASE}     & Recall                  & 52.11                  & 38.61$\rpm$0.42 & 44.85$\rpm$1.33 & 48.31$\rpm$2.77 & 48.61$\rpm$2.79 & 48.79$\rpm$3.85  \\ \cline{2-8} 
                          & NDCG                    & 26.19                  & 17.67$\rpm$0.20 & 20.59$\rpm$0.64 & 22.29$\rpm$1.38 & 22.43$\rpm$1.39 & 22.55$\rpm$1.95  \\ \cline{2-8} 
                          & AP                      & 18.46                  & 11.46$\rpm$0.13 & 13.40$\rpm$0.43 & 14.57$\rpm$0.96 & 14.66$\rpm$0.97 & 14.76$\rpm$1.38  \\ \hline
\multirow{3}{*}{MultiVAE} & Recall                  & 24.63                  & 36.13$\rpm$0.48 & 37.46$\rpm$1.30 & 36.01$\rpm$2.16 & 36.14$\rpm$2.17 & 34.28$\rpm$3.01  \\ \cline{2-8} 
                          & NDCG                    & 9.72                   & 16.49$\rpm$0.23 & 17.07$\rpm$0.62 & 16.34$\rpm$1.04 & 16.40$\rpm$1.05 & 15.49$\rpm$1.47  \\ \cline{2-8} 
                          & AP                      & 5.39                   & 10.68$\rpm$0.15 & 11.04$\rpm$0.42 & 10.52$\rpm$0.71 & 10.55$\rpm$0.71 & 9.93$\rpm$1.00   \\ \hline
\multirow{3}{*}{NeuMF}    & Recall                  & 41.91                  & 34.82$\rpm$0.51 & 38.94$\rpm$1.36 & 40.55$\rpm$2.45 & 40.74$\rpm$2.46 & 40.40$\rpm$3.58  \\ \cline{2-8} 
                          & NDCG                    & 19.75                  & 15.92$\rpm$0.24 & 17.83$\rpm$0.65 & 18.61$\rpm$1.21 & 18.70$\rpm$1.21 & 18.55$\rpm$1.80  \\ \cline{2-8} 
                          & AP                      & 13.14                  & 10.32$\rpm$0.16 & 11.58$\rpm$0.44 & 12.11$\rpm$0.83 & 12.16$\rpm$0.84 & 12.07$\rpm$1.26  \\ \hline
\multirow{3}{*}{itemKNN}  & Recall                  & 52.53                  & 39.04$\rpm$0.42 & 45.30$\rpm$1.21 & 48.68$\rpm$2.47 & 48.91$\rpm$2.49 & 48.96$\rpm$3.54  \\ \cline{2-8} 
                          & NDCG                    & 26.23                  & 17.87$\rpm$0.20 & 20.79$\rpm$0.58 & 22.45$\rpm$1.23 & 22.56$\rpm$1.24 & 22.61$\rpm$1.79  \\ \cline{2-8} 
                          & AP                      & 18.38                  & 11.59$\rpm$0.13 & 13.53$\rpm$0.39 & 14.67$\rpm$0.86 & 14.74$\rpm$0.86 & 14.79$\rpm$1.26  \\ \hline
\multirow{3}{*}{ALS}      & Recall                  & 43.40                  & 34.34$\rpm$0.43 & 39.26$\rpm$1.17 & 41.77$\rpm$2.24 & 42.32$\rpm$2.27 & 41.90$\rpm$3.21  \\ \cline{2-8} 
                          & NDCG                    & 20.62                  & 15.71$\rpm$0.20 & 18.01$\rpm$0.56 & 19.23$\rpm$1.11 & 19.48$\rpm$1.13 & 19.31$\rpm$1.62  \\ \cline{2-8} 
                          & AP                      & 13.83                  & 10.19$\rpm$0.13 & 11.71$\rpm$0.38 & 12.54$\rpm$0.77 & 12.71$\rpm$0.78 & 12.61$\rpm$1.14  \\ \hline
\end{tabular}
\caption{Dataset: pinterest-20 with sample size =$99$.}
\label{tab:pinterest_base}
\end{table*}
\end{small}

% \usepackage{multirow}
\begin{small}
\begin{table*}[t]
\begin{tabular}{|c|c|c|c|c|c|c|c|}
\hline
\multirow{2}{*}{Model}    & \multirow{2}{*}{Metric} & \multirow{2}{*}{Exact} & \multicolumn{5}{c|}{Estimators of $Metric@10$}                                   \\ \cline{4-8} 
                          &                         &                        & BV 0.1        & BV 0.01       & MLE           & WMLE         & MES \\ \hline
\multirow{3}{*}{EASE}     & Recall                  & 25.548                 & 17.00$\rpm$0.26 & 21.30$\rpm$0.81 & 25.31$\rpm$2.24 & 25.78$\rpm$2.28 & 25.59$\rpm$2.87  \\ \cline{2-8} 
                          & NDCG                    & 11.848                 & 7.75$\rpm$0.12  & 9.73$\rpm$0.38  & 11.62$\rpm$1.06 & 11.83$\rpm$1.08 & 11.75$\rpm$1.36  \\ \cline{2-8} 
                          & AP                      & 7.794                  & 5.01$\rpm$0.08  & 6.30$\rpm$0.25  & 7.56$\rpm$0.71  & 7.69$\rpm$0.72  & 7.65$\rpm$0.91   \\ \hline
\multirow{3}{*}{MultiVAE} & Recall                  & 15.383                 & 15.57$\rpm$0.25 & 18.79$\rpm$0.73 & 21.05$\rpm$1.81 & 21.38$\rpm$1.84 & 21.21$\rpm$2.26  \\ \cline{2-8} 
                          & NDCG                    & 5.741                  & 7.10$\rpm$0.11  & 8.57$\rpm$0.34  & 9.63$\rpm$0.85  & 9.78$\rpm$0.87  & 9.71$\rpm$1.07   \\ \cline{2-8} 
                          & AP                      & 2.943                  & 4.59$\rpm$0.07  & 5.55$\rpm$0.22  & 6.25$\rpm$0.57  & 6.35$\rpm$0.58  & 6.30$\rpm$0.71   \\ \hline
\multirow{3}{*}{NeuMF}    & Recall                  & 21.381                 & 13.84$\rpm$0.24 & 16.61$\rpm$0.69 & 18.56$\rpm$1.62 & 18.86$\rpm$1.64 & 18.61$\rpm$2.00  \\ \cline{2-8} 
                          & NDCG                    & 10.094                 & 6.31$\rpm$0.11  & 7.58$\rpm$0.32  & 8.49$\rpm$0.76  & 8.63$\rpm$0.77  & 8.52$\rpm$0.94   \\ \cline{2-8} 
                          & AP                      & 6.739                  & 4.08$\rpm$0.07  & 4.91$\rpm$0.21  & 5.51$\rpm$0.51  & 5.60$\rpm$0.51  & 5.53$\rpm$0.63   \\ \hline
\multirow{3}{*}{itemKNN}  & Recall                  & 35.129                 & 19.90$\rpm$0.31 & 25.68$\rpm$0.98 & 31.85$\rpm$3.02 & 32.40$\rpm$3.08 & 32.51$\rpm$3.79  \\ \cline{2-8} 
                          & NDCG                    & 16.258                 & 9.08$\rpm$0.14  & 11.74$\rpm$0.45 & 14.65$\rpm$1.43 & 14.90$\rpm$1.46 & 14.97$\rpm$1.81  \\ \cline{2-8} 
                          & AP                      & 10.670                 & 5.87$\rpm$0.09  & 7.61$\rpm$0.30  & 9.55$\rpm$0.96  & 9.71$\rpm$0.98  & 9.76$\rpm$1.22   \\ \hline
\multirow{3}{*}{ALS}      & Recall                  & 20.057                 & 15.04$\rpm$0.28 & 18.60$\rpm$0.80 & 21.77$\rpm$2.08 & 22.56$\rpm$2.16 & 22.02$\rpm$2.41  \\ \cline{2-8} 
                          & NDCG                    & 8.545                  & 6.86$\rpm$0.13  & 8.49$\rpm$0.37  & 9.98$\rpm$0.98  & 10.35$\rpm$1.02 & 10.11$\rpm$1.14  \\ \cline{2-8} 
                          & AP                      & 5.131                  & 4.43$\rpm$0.08  & 5.50$\rpm$0.24  & 6.49$\rpm$0.65  & 6.73$\rpm$0.68  & 6.57$\rpm$0.76   \\ \hline
\end{tabular}
\caption{Dataset: yelp with sample size =$99$.}
\label{tab:yelp_base}
\end{table*}
\end{small}

\section{Predicting Winners by Metric@$K$}
Table~\ref{tab:winner_citeulike},~\ref{tab:winner_yelp} and ~\ref{tab:winner_pinterest}  show, among the $100$ sample runs, the number of correct winners (algorithm with best performance) predicted by $BV$, $MLE$, $WLE$ and $MES$ estimators based on $Recall@K$, $NDCG@K$ and $AP@K$ for $K=1$, $5$, $10$ and $20$, on the $pinterest-20$, $yelp$ and $citeulike$ dataset respectively.

% \usepackage{multirow}
\begin{table}[]
\begin{tabular}{|c|c|l|l|l|l|l|}
\hline
K                   & Metric & \multicolumn{1}{c|}{BV0.1} & \multicolumn{1}{c|}{BV0.01} & \multicolumn{1}{c|}{MLE} & \multicolumn{1}{c|}{WMLE} & \multicolumn{1}{c|}{MES} \\ \hline
\multirow{3}{*}{1}  & Recall & 0                          & 5                           & 22                      & 22                         & 28                                  \\ \cline{2-7} 
                    & NDCG   & 0                          & 5                           & 22                      & 22                         & 28                                  \\ \cline{2-7} 
                    & AP     & 0                          & 5                           & 22                      & 22                         & 28                                  \\ \hline
\multirow{3}{*}{5}  & Recall & 0                          & 5                           & 21                      & 21                         & 27                                  \\ \cline{2-7} 
                    & NDCG   & 0                          & 5                           & 21                      & 22                         & 28                                  \\ \cline{2-7} 
                    & AP     & 0                          & 5                           & 21                      & 22                         & 28                                  \\ \hline
\multirow{3}{*}{10} & Recall & 0                          & 5                           & 18                      & 19                         & 26                                  \\ \cline{2-7} 
                    & NDCG   & 0                          & 5                           & 19                      & 19                         & 26                                  \\ \cline{2-7} 
                    & AP     & 0                          & 5                           & 20                      & 21                         & 27                                  \\ \hline
\multirow{3}{*}{20} & Recall & 0                          & 3                           & 12                      & 13                         & 19                                  \\ \cline{2-7} 
                    & NDCG   & 0                          & 4                           & 15                      & 16                         & 25                                  \\ \cline{2-7} 
                    & AP     & 0                          & 5                           & 18                      & 19                         & 26                                  \\ \hline
\end{tabular}
\caption{The number of successes at predicting a winner on the citeulike dataset with 100 repeats.}
\label{tab:winner_citeulike}
\vspace{-3.0ex}
\end{table}

% \usepackage{multirow}
\begin{table}[]
\begin{tabular}{|c|c|c|c|c|c|c|}
\hline
K                   & Metric & BV0.1 & BV0.01 & MLE & WMLE & MES \\ \hline
\multirow{3}{*}{1}  & Recall & 100   & 100    & 97 & 97    & 94             \\ \cline{2-7} 
                    & NDCG   & 100   & 100    & 97 & 97    & 94             \\ \cline{2-7} 
                    & AP     & 100   & 100    & 97 & 97    & 94             \\ \hline
\multirow{3}{*}{5}  & Recall & 100   & 100    & 97 & 97    & 94             \\ \cline{2-7} 
                    & NDCG   & 100   & 100    & 97 & 97    & 94             \\ \cline{2-7} 
                    & AP     & 100   & 100    & 97 & 97    & 94             \\ \hline
\multirow{3}{*}{10} & Recall & 100   & 100    & 97 & 97    & 94             \\ \cline{2-7} 
                    & NDCG   & 100   & 100    & 97 & 97    & 94             \\ \cline{2-7} 
                    & AP     & 100   & 100    & 97 & 97    & 94             \\ \hline
\multirow{3}{*}{20} & Recall & 100   & 100    & 99 & 99    & 97             \\ \cline{2-7} 
                    & NDCG   & 100   & 100    & 99 & 99    & 96             \\ \cline{2-7} 
                    & AP     & 100   & 100    & 97 & 97    & 94             \\ \hline
\end{tabular}
\caption{The number of successes at predicting a winner on the yelp dataset with 100 repeats.}
\label{tab:winner_yelp}
\vspace{-3.0ex}
\end{table}

% \usepackage{multirow}
\begin{table}[]
\begin{tabular}{|c|c|c|c|c|c|c|}
\hline
K                   & Metric & BV0.1 & BV0.01 & MLE & WMLE & MES \\ \hline
\multirow{3}{*}{1}  & Recall & 17    & 36     & 46 & 46    & 48             \\ \cline{2-7} 
                    & NDCG   & 17    & 36     & 46 & 46    & 48             \\ \cline{2-7} 
                    & AP     & 17    & 36     & 46 & 46    & 48             \\ \hline
\multirow{3}{*}{5}  & Recall & 86    & 64     & 55 & 55    & 48             \\ \cline{2-7} 
                    & NDCG   & 16    & 36     & 45 & 45    & 49             \\ \cline{2-7} 
                    & AP     & 16    & 36     & 45 & 45    & 49             \\ \hline
\multirow{3}{*}{10} & Recall & 86    & 66     & 57 & 55    & 51             \\ \cline{2-7} 
                    & NDCG   & 86    & 65     & 57 & 53    & 50             \\ \cline{2-7} 
                    & AP     & 14    & 36     & 45 & 45    & 48             \\ \hline
\multirow{3}{*}{20} & Recall & 87    & 69     & 61 & 59    & 56             \\ \cline{2-7} 
                    & NDCG   & 87    & 66     & 59 & 56    & 52             \\ \cline{2-7} 
                    & AP     & 86    & 66     & 56 & 55    & 50             \\ \hline
\end{tabular}
\caption{The number of successes at predicting a winner on the pinterest-20 dataset with 100 repeats.}
\label{tab:winner_pinterest}
\vspace{-3.0ex}
\end{table}
\end{document}